\documentclass{aa}

\usepackage{graphics}

\begin{document}

\thesaurus{03           
             (03.13.2;  
              11.09.1;  
              11.16.2;  
              11.19.4)} 

\title{Globular clusters in NGC 5128\thanks{Based in observations with
the NASA/ESA \emph{Hubble Space Telescope}, obtained at the Space
Telescope Science Institute, which is operated by the Association of
Universities for Research in Astronomy, Inc., under NASA contract
NAS5-26555.}}

\subtitle{}

\author{S. Holland\inst{1}
             \and
        P. C{\^o}t{\'e}\inst{2}\thanks{Sherman M. Fairchild Fellow}
             \and
        J. E. Hesser\inst{3}}

\offprints{S. Holland}

\institute{Institut for Fysik og Astronomi (IFA),
           Aarhus Universitet,
           Ny Munkegade, Bygning 520,
           DK--8000 {\AA}rhus C,
           Denmark,
           e-mail: holland@obs.aau.dk
                \and
           California Institute of Technology,
           Mail Stop 105--24,
           Pasadena, California, 91125
           USA,
           e-mail: pc@astro.caltech.edu
                \and
           Dominion Astrophysical Observatory,
           Herzberg Institute of Astrophysics,
           National Research Council,
           5071 West Saanich Road,
           Victoria, B.C., V8X 4M6,
           Canada,
           e-mail: Jim.Hesser@hia.nrc.ca}

\date{Received 03 February 1999; accepted}

\maketitle


\begin{abstract}

	We used the Wide Field Planetary Camera 2 aboard the
\emph{Hubble Space Telescope} to search for globular clusters in the
inner regions of the nearby giant elliptical galaxy \object{NGC 5128}.
This galaxy is believed to be the product of a merger between a large
elliptical galaxy and a small late-type spiral between 160 and 500 Myr
ago.  We identified 21 globular cluster candidates and measured their
core radii, tidal radii, half-mass radii, ellipticities, position
angles, and $V\!-\!I$ colors.  We find evidence that the \object{NGC
5128} globular cluster candidates are systematically more elliptical
than are those of the \object{Milky Way}.  Approximately half of the
candidates have ${(V\!-\!I)}_0$ colors that are consistent with their
being either old, unreddened globular clusters, similar to those found
in the \object{Milky Way}, or young, reddened globular clusters that
may have formed during the recent merger event.  Most of the rest have
colors that are consistent with their being old globular clusters
similar to those found in the \object{Milky Way}.  We find one blue
object with ${(V\!-\!I)}_0 < 0.26 \pm 0.09$.  The color, reddening,
and integrated magnitude of this object are consistent with its being
a small globular cluster with an age of $\sim 100$ Myr and a mass
(based on its integrated luminosity) of $\le 4000$ M$_\odot$.  We find
no evidence for bimodality in the colors of the globular cluster
candidates in our sample beyond what can be explained by uncertainties
in the differential reddening.


\end{abstract}

%
%

\section{Introduction\label{SECTION:intro}}

       Globular star clusters (GCs) are among the oldest stellar
systems in the Universe and provide a powerful tracer of the
evolutionary history of a galaxy.  There is strong evidence that
massive star clusters can form during galactic mergers (e.g. Zepf \&
Ashman~\cite{ZA93}; Schweizer et~al.~\cite{ScM96}), so galaxies that
have recently experienced a merger event are ideal places to search
for young GCs.  Recently several candidates for luminous young GCs
have been identified in various merging galaxies, such as \object{NGC
3597} (Lutz~\cite{L91}), \object{NGC 1275} (Holtzman
et~al.~\cite{HF92}), and \object{NGC 7252} (Schweizer \&
Seitzer~\cite{SS93}).

         \object{NGC 5128} ($=$ \object{Centaurus A}, see
Israel~\cite{I98} for a recent review) is classified as a giant S0pec
galaxy.  It is composed of a large, dominant spheroid, which itself
resembles an E0 galaxy, and a disk that contains large amounts of gas
and dust.  Soria et~al.~(\cite{SM96}) used direct observations of
resolved stars at the tip of the red giant branch in \object{NGC 5128}
to determine a true distance modulus of $\mu_0 = 27.8 \pm 0.2$, which
corresponds to $3.6 \pm 0.2$ Mpc, making \object{NGC 5128} the nearest
giant elliptical galaxy to our own.  There is strong evidence (see the
review by Ebneter \& Balick~\cite{EB83}) that it is the product of a
recent merger between a large elliptical galaxy and a small late-type
spiral.  A thick dust band is seen across the center of \object{NGC
5128} and there is evidence for significant star formation occurring
in the central regions of the galaxy.  G. Harris et~al.~(\cite{HH99})
used \emph{Hubble Space Telescope} (\emph{HST}) Wide Field Planetary
Camera 2 (WFPC2) images to obtain a color--magnitude diagram for the
outer halo of \object{NGC 5128}.  They found a distance of $3.9 \pm
0.3$ Mpc, consistent with the Soria et~al.~(\cite{SM96}) estimate, and
a population of old stars with iron abundances between
$\mathrm{[Fe/H]} \sim -2$ and $\mathrm{[Fe/H]} \sim +0.2$.  Their
metallicity distribution function is consistent with two bursts of
star formation.  The first having $\mathrm{[m/H]} = -0.6$ and
producing approximately one-third of the stars, and the second having
$\mathrm{[m/H]} = 0$ and producing approximately two-thirds of the
stars.  They argue that the second burst of star formation occurred at
least 1--2 Gyr after the first.

        The first observation of a GC in \object{NGC 5128} was by
Graham \& Philips~(\cite{GP80}).  The galaxy is now believed to have
$\sim 1700$ GCs (H. Harris et~al.~\cite{HH84}), with 87 confirmed
spectroscopically (see H. Harris et~al.~\cite{HH88};
Sharples~\cite{S88}).  Recently G. Harris et~al.~(\cite{HP98}) used
\emph{HST} WFPC2 images to construct a color--magnitude diagram for
\object{C44}, a GC in the halo of \object{NGC 5128}.  They found that
this GC was an old, intermediate-metallicity object similar to the GCs
in the \object{Milky Way}.  G. Harris et~al.~(\cite{HG92}, hereafter
referred to as HG92) used Washington $CM{T_1}{T_2}$ photometry to
derive metallicities for 62 of confirmed GCs in \object{NGC 5128} and
found a mean iron abundance of $\mathrm{[Fe/H]} = -0.8 \pm 0.2$, which
suggest that the \object{NGC 5128} GC system is $\sim 3$ times more
metal rich than the \object{Milky Way} GC system.  They found no
evidence for any GCs having metallicities significantly greater than
those found in the \object{Milky Way} GCs.  Such metal-rich GCs might
be expected if some of the \object{NGC 5128} GCs had formed recently
in a gas-rich merger event.  HG92 do, however, suggest that several
blue GCs in \object{NGC 5128} may be analogues of the intermediate-age
GCs found in the Magellanic Clouds. On the other hand, Zepf \&
Ashman~(\cite{ZA93}) suggest that the metallicity distribution of the
\object{NGC 5128} GCs is bimodal, with the high-metallicity peak at
$\mathrm{[Fe/H]} = +0.25$ due to GCs formed in a merger.  Hui
et~al.~(\cite{HF95}) analyzed the kinematics of the \object{NGC 5128}
GC system and found that the metal-rich GCs are part of a dynamically
separate system from the metal-poor GCs.  Numerical simulations
suggest that the merger event occurred between 160 (Quillen
et~al.~\cite{QG93}) and 500 ($D/5$ Mpc) Myr ago, where $D$ is the
distance to \object{NGC 5128} in Mpc (Tubbs~\cite{T80}).  This
suggests that any GCs that formed in \emph{this particular} merger
should be quite young and, therefore, rather blue ($\sim 0.4 < V\!-\!I
< 0.6$; see Sect.~\ref{SECTION:young_GCs}).

	Minniti et~al.~(\cite{MA96}) and Alonso \&
Minniti~(\cite{AM97}, hereafter referred to as AM97) used \emph{HST}
Wide-Field/Planetary-Camera 1 (WF/PC-1) images, taken before the
corrective optics package was installed in 1993, to search for GCs in
the inner regions of \object{NGC 5128}.  They identified 125 GC
candidates, young associations, and open cluster candidates in the
inner three kpc of \object{NGC 5128}.  They also used ground-based
$RK$ photometry to estimate metallicities for 47 GC candidates.
Schreier et~al.~(\cite{SC96}) found 74 compact sources along the
northern edge of the \object{NGC 5128} dust lane using \emph{HST}
WF/PC-1 images.  They estimate that most of these sources are young
stars (spectral class A or earlier) but note that some are resolved
and may be GCs.

        Identifying GC candidates in the inner regions of \object{NGC
5128} is difficult since there is nonuniform extinction, contamination
from foreground stars and background galaxies, and confusion with open
clusters and blue, star-forming knots in \object{NGC 5128}.  GC
candidates can not be identified based solely on their colors since
the large amount of uneven reddening makes it very difficult to
determine the dereddened color of an object.  A better approach is a
scheme to identify GC candidates based solely on their structural
parameters.  All known GCs in the Local Group can be reasonably well
fit by Michie--King models (Michie~\cite{M63}; King~\cite{K66}),
although $\sim 20$\% show evidence of having undergone core collapse.
The vast majority of the Milky Way's GCs are uniformly old objects
with ages of $11.5 \pm 1.3$ Gyr (Chaboyer et~al.~\cite{CD98}), mean
King core radii of $r_c = 2.3 \pm 0.4$ pc, mean King tidal radii of
$r_t = 45.2 \pm 3.5$ pc, mean concentrations of $c \equiv
\log_{10}(r_t/r_c) = 1.40 \pm 0.04$, and mean ellipticities of
$\epsilon = 0.08 \pm 0.01$.  If the GCs in \object{NGC 5128} are
structurally similar to those in the Local Group spiral and dwarf
galaxies, then high resolution imaging can be used to identify GC
candidates in the inner regions of \object{NGC 5128} without resorting
to an identification scheme based upon the integrated colors of the
objects.


\section{The Data\label{SECTION:data}}

\subsection{Observations\label{SECTION:obs}}

        We used the WFPC2 aboard the \emph{HST} to obtain F555W (WFPC2
broadband $V$) and F814W (WFPC2 broadband $I$) images of a region near
the nucleus of \object{NGC 5128}.  The data were obtained on July 27,
1997 for the cycle 6 program GO-6789.  The WFPC2 had an operating
temperature of $-88\degr$C and a nominal gain setting of 7
$\mathrm{e}^-$/ADU\@.  The observations are listed in
Table~\ref{TABLE:obs_log}.


%
%

\begin{table}
\begin{center}
\caption[]{Log of the observations.}
\smallskip
\begin{tabular}{ccccc}
\hline
\hline
\vspace{0.1 em}
Field & $\alpha$(J2000) & $\delta$(J2000) & Filter & Exposure \\
\hline
\vspace{0.1 em}
1 & $13^\mathrm{h}25^\mathrm{m}33\fs5$ & $-43\degr00\arcmin14\arcsec$ & F555W & 3 $\times$ 60 \\
  &                                    &                              & F814W & 3 $\times$ 60 \\
2 & $13^\mathrm{h}25^\mathrm{m}28\fs1$ & $-43\degr00\arcmin14\arcsec$ & F555W & 4 $\times$ 60 \\
  &                                    &                              & F814W & 3 $\times$ 60 \\
3 & $13^\mathrm{h}25^\mathrm{m}22\fs7$ & $-43\degr00\arcmin14\arcsec$ & F555W & 3 $\times$ 60 \\
  &                                    &                              & F814W & 3 $\times$ 60 \\
4 & $13^\mathrm{h}25^\mathrm{m}33\fs6$ & $-43\degr02\arcmin14\arcsec$ & F555W & 3 $\times$ 40 \\
  &                                    &                              & F814W & 3 $\times$ 40 \\
5 & $13^\mathrm{h}25^\mathrm{m}28\fs2$ & $-43\degr02\arcmin14\arcsec$ & F555W & 3 $\times$ 60 \\
  &                                    &                              & F814W & 3 $\times$ 50 \\
6 & $13^\mathrm{h}25^\mathrm{m}22\fs7$ & $-43\degr02\arcmin14\arcsec$ & F555W & 3 $\times$ 60 \\
  &                                    &                              & F814W & 3 $\times$ 60 \\
\hline
\hline
\end{tabular}
\label{TABLE:obs_log}
\end{center}
\end{table}

        Exposures were taken in each field with each of the F555W and
the F814W filters.  Cosmic rays impact $\sim 20$ pixels per second on
each WFPC2 CCD, but by combining the multiple exposures per filter for
each field, the number of pixels lost to cosmic ray events is
negligible.  Therefore, we did not apply any processing explicitly to
remove cosmic rays from the images.  The data were preprocessed
through the standard STScI pipeline for WFPC2 data.  Known bad pixels
were flagged and not used in the data analysis.  No corrections were
made for geometric distortions in the area of the WFPC2 pixels.

        The survey consists of six fields that cover a total area of
approximately $5\farcm3 \times 8\farcm0$ centered on $\alpha =
13^\mathrm{h}25^\mathrm{m}27{\fs}3, \delta =
-43{\degr}01{\arcmin}09{\arcsec}$ (J2000 coordinates), the nucleus of
\object{NGC 5128}.  Adjacent fields overlap by $\sim 0\farcm5$ giving
a total effective area of $\sim 25$ $\sq\arcmin$ for the survey.

\subsection{Data Reductions\label{SECTION:data_reductions}}

        We combined the exposures for each field by taking the average
of the three images in each filter (four images for the F555W
exposures of Field 2).  No re-registration of the images was performed
since the shifts between the images were typically less than 0.1 pixel
($0\farcs01$ on the WFC and $0\farcs005$ on the PC).  We estimate that
combining the images in this way may result in the sizes of the GC
candidates being systematically overestimated by no more than $\sim
0\farcs02$.  We prefer to introduce this simple systematic offset than
deal with the poorly-understood systematic uncertainties that arise
from interpolating flux across fractional-pixel shifts.

\subsubsection{Identifying Globular Cluster Candidates\label{SECTION:find}}

	At the distance of \object{NGC 5128} ($d = 3.6 \pm 0.2$ Mpc),
the mean King core- and tidal-radii of the Milky Way GCs would appear
to be $\overline{r_c} = 0\farcs13 \pm 0\farcs02$ and $\overline{r_t} =
2\farcs59 \pm 0\farcs20$, respectively.  Therefore, any GCs in
\object{NGC 5128} will appear to be semi-stellar and be strongly
affected by the point spread function (PSF) of the WFPC2.  After some
experimentation, we adopted the following procedure for identifying GC
candidates.  We wish to stress that this procedure is quite strict and
will probably result in the rejection of some legitimate GC
candidates.  However, we prefer to reject real GCs rather than have
our sample contaminated with stars or background galaxies.

        In order to increase the signal-to-noise ratio (S/N) of the GC
candidates -- a particularly important point for the faint ($V < 20$)
GC candidates -- we combined the F555W and F814W images for each field
to get finding images.  The dust lane introduces variations in the
background on spatial scales of $\sim 1\arcsec$, comparable to the
expected sizes of the GC candidates in \object{NGC 5128}.  To reduce
the effects of the uneven background light, large-scale spatial
variations in the background were removed by running a ring median
filter (Secker~\cite{S95}) over the finding image, subtracting the
resulting smoothed background, and adding back the mean background
value.  The median filter radius was set to $1\arcsec$, which is $\sim
3.5$ times the expected full-width at half maximum (FWHM) of a typical
GC candidate.  This choice of filter radius ensures that the cores of
the GC candidates will not be altered by the median filter and that
any background structure larger than a typical GC candidate will be
removed.  Since the most extended Milky Way GCs have tidal radii that
are significantly greater than 2.5 times their FWHM, and extended
halos have been detected around several Galactic and extra-Galactic
GCs (Grillmair et~al.~\cite{GF95};~\cite{GA96}; Holland
et~al.~\cite{HF97}), this approach will alter the distribution of
light in the outer regions of most of the GC candidates.  However,
this is not important since the finding images are used \emph{only} to
construct a preliminary list of GC candidates.  A more rigorous set of
criteria, based on the structures of the GC candidates as determined
from the \emph{original} images, will be applied to the preliminary
list to obtain a final list of GC candidates in the central regions of
\object{NGC 5128}.

        The first step in our identification procedure was to run the
{\sc daophot ii} (Stetson~\cite{S87};~\cite{S94}) {\sc find} routine
on the background-subtracted images to identify GC candidates.  The
finding thresholds were set to $6\sigma_\mathrm{sky}$ for the PC
images and $10\sigma_\mathrm{sky}$ for the WFC images.  Tests with
artificial GCs suggested that any detections below these thresholds
would be rejected at some point in our identification process.  {\sc
Daophot ii find} has an algorithm for rejecting non-stellar objects
based on two parameters called ``sharpness'' and ``round''.  This
algorithm was turned off since images of GCs can have different shapes
and concentrations from images of stars.

        Next, the {\sc daophot ii photometry} routine was used to
obtain aperture photometry for each of these detections.  The
photometry was performed separately on each of the combined F555W and
F814W frames, \emph{not} on the combined finding frame.  An aperture
radius of $0\farcs2$ was used since most Galactic GCs, if moved to the
distance of \object{NGC 5128}, would appear to have core radii smaller
than this.  Therefore, the signal within the aperture will be
dominated by the light from the object and not from the background.
Candidate objects with $\mathrm{S/N} < 5$ within the photometry
aperture were discarded since the signal was not strong enough to
determine reliable shape parameters.  The sky brightness was
determined in an annulus with an inner radius of $\sim 0\farcs9$ and
an outer radius of $\sim 1\farcs1$.  This annulus was chosen to be far
enough from the center of the GC candidate that the light in the
annulus will be dominated by the background, yet near enough to the GC
candidate that the light in the annulus will be a reasonable
approximation of the mean background at the location of the object.
For large GC candidates this annulus will be inside the tidal radius
of the object so our estimate of the background will be contaminated.
However, the values determined at this stage are only preliminary
estimates, which will be improved upon later in the identification
process when Michie--King models are fit to the GC candidates.  The
lists of GC candidates in each of the F555W and F814W images were
matched using the {\sc daomatch} and {\sc daomaster} software.  Only
objects that appeared in \emph{both} the F555W and F814W images, and
whose centers matched to within $0\farcs05$ ($\sim 1.1$ pixel on the
PC images and $\sim 0.5$ pix on the WFC images), were considered to be
real GC candidates.

        Distinguishing bona fide GCs from stars and background
galaxies is challenging.  The colors of the objects can not be used
since we are interested in studying the color distribution of GCs in
\object{NGC 5128} and do not wish to bias our sample.  To make matters
worse, the presence of dust in \object{NGC 5128} will add a
significant amount of scatter to the intrinsic color distribution, and
may cause legitimate GCs to be rejected if a color-based
identification scheme is used.  The solution is to identify GC
candidates by their structural parameters, although the best choice of
structural parameters is not obvious.  At the distance of \object{NGC
5128} a typical Galactic GC would appear to have an intrinsic FWHM of
$\sim 0\farcs25$, or approximately twice the FWHM of the WFPC2 PSFs.
Therefore, the observed FWHM, concentration, and ellipticity of a GC
candidate can be heavily influenced by the PSF\@.  Since the PSF
varies strongly with position on the WFPC2 CCDs, the potential for
confusion between stellar images and concentrated GC candidates is
great if the PSF is not removed, in some way, from the data.
Therefore, the observed shape of an object can not be directly used to
classify it as a star, GC candidate, or galaxy.

        After some experimentation with adding and recovering
artificial GCs and artificial stars, we found that the following
procedure was reasonably reliable for identifying GC candidates.  For
each GC candidate we took all the pixel values within $1\arcsec$ of
the center of the object and subtracted an estimate of the local
background (the center and background were determined by the {\sc
daophot ii photometry} algorithm).  A one-dimensional Moffatian
(Moffat~\cite{M69}),

\begin{equation}
M(r_\mathrm{eff}) = M(0) {\left[ 1 + {\left(r_\mathrm{eff}
                      \over \alpha\right)}^2 \right]}^{-\beta},
\label{EQUATION:moffat}
\end{equation}

\noindent
was fit to each candidate using the effective radius,
$r_\mathrm{eff}$, instead of the true distance from the center of the
candidate in order to compensate for any ellipticity that might be
introduced by the PSF\@.  The effective radius of an ellipse is
defined by $a b \pi \equiv r_\mathrm{eff}^2 \pi$, where $a$ and $b$
are the lengths of the semi-major and semi-minor axes of the ellipse,
respectively.  For each pixel the effective radius from the center of
the GC candidate was computed using:

\begin{eqnarray}
r_\mathrm{eff}^2 & = & x^2\left[{(1-\epsilon)}^2 \cos^2\theta_0
           + \sin^2\theta_0\right] / (1-\epsilon) \nonumber \\
                 &   & \mbox{} + y^2\left[\cos^2\theta_0
           + {(1-\epsilon)}^2 \sin^2\theta_0\right] / (1-\epsilon) \nonumber \\
                 &   & \mbox{} + 2xy\epsilon(\epsilon-2)
                   \cos\theta_0\sin\theta_0 / (1-\epsilon),
\label{EQUATION:r_eff}
\end{eqnarray}

\noindent
where $x$ and $y$ are the coordinates of the pixel on the CCD, and
$\epsilon$ and $\theta_0$ are estimates of the ellipticity and
position angle of the GC candidate.  The latter two quantities were
estimated from the moments of the light from each object.

\begin{figure}
\resizebox{\hsize}{!}{\includegraphics{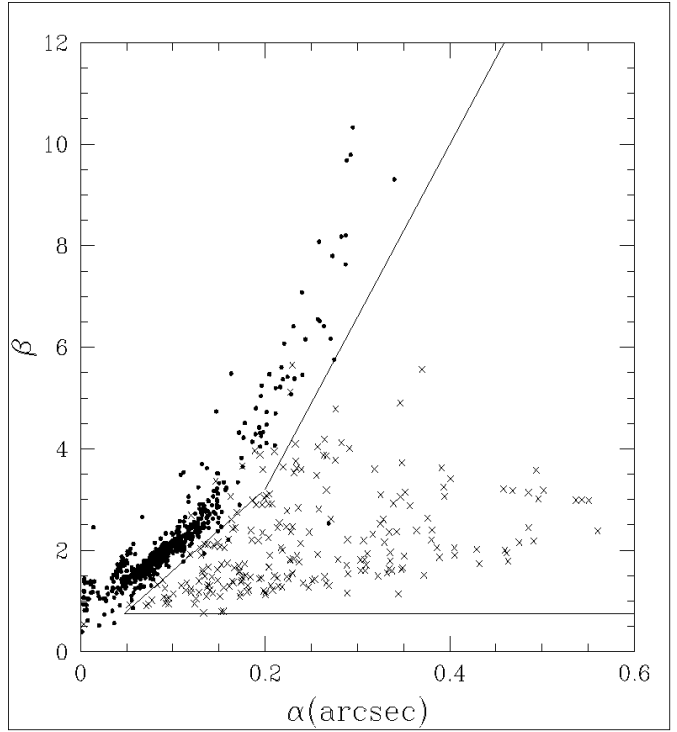}}
\caption{This figure shows the best-fitting Moffatian $\alpha$ and
$\beta$ parameters for the artificial stars (circles) and artificial
GCs (crosses).  Based on the distribution of objects in this diagram
we assumed that any objects that lie inside the wedge formed by the
solid lines were GC candidates.  No upper limit was placed on the
value of $\beta$.}
\label{FIGURE:moffat_art}
\end{figure}

	In order to determine which combinations of $\alpha$ and
$\beta$ corresponded to stars and which corresponded to GC candidates,
a series of artificial stars and artificial GCs were added to the
images.  The artificial stars were added using the {\sc daophot ii
addstar} routine and the appropriate PSFs scaled to magnitudes of $16
\le \mathrm{F555W} \le 22$.  The artificial GCs, also with integrated
magnitudes of $16 \le \mathrm{F555W} \le 22$, were added using the
IRAF\footnote{Image Reduction and Analysis Facility (IRAF), a software
system distributed by the National Optical Astronomy Observatories
(NOAO).} v2.10.4 task {\sc noao.artdata.mkobject}.  The artificial GCs
all had ellipticities of $\epsilon = 0$, concentrations of $0.67 \le c
\le 2.12$, and core radii of $0\farcs067 \le r_c \le 0\farcs4$
(corresponding to physical core radii of between $\sim 1$ and $\sim 7$
pc at the distance of \object{NGC 5128}).  Therefore, the artificial
GCs had a range of structures similar to those of the \object{Milky
Way}'s GCs.  The procedure described above was used to determine the
Moffat $\alpha$ and $\beta$ for each artificial object.  The results
are presented in Fig.~\ref{FIGURE:moffat_art} and were used to
determine which combinations of $\alpha$ and $\beta$ represent stars
and which represent GC candidates.  These limits on $\alpha$ and
$\beta$ were then applied to the GC candidates found on the WFPC2
images of \object{NGC 5128}.

\begin{figure}
\resizebox{\hsize}{!}{\includegraphics{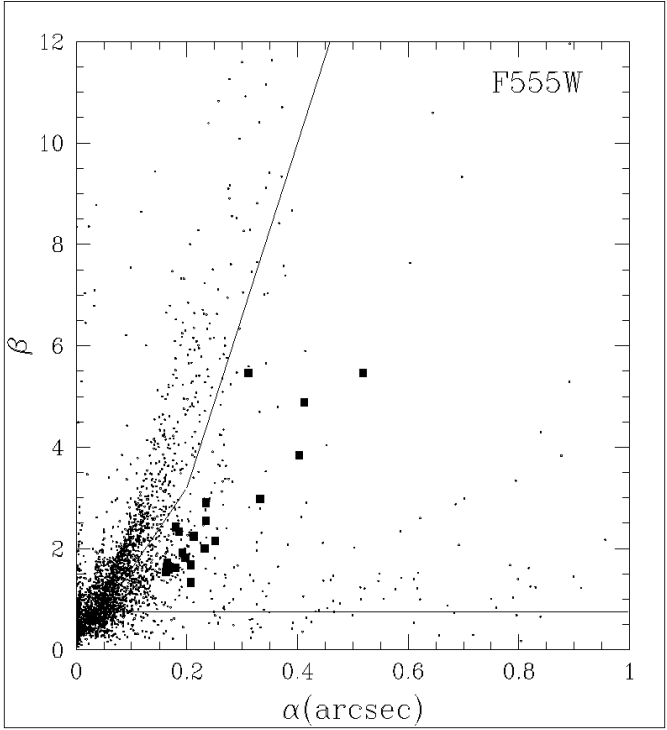}}
\caption{This figure shows the best-fitting Moffatian $\alpha$ and
$\beta$ parameters for each object (small circles) on the F555W
images.  Simulated data (see Fig.~\ref{FIGURE:moffat_art}) suggest
that objects that lie inside the wedge formed by the solid lines are
extended objects.  Therefore we consider any objects that lie inside
the wedge to to be GC candidates.  The solid squares show the
locations of the GC candidates from Table~\ref{TABLE:objects}.}
\label{FIGURE:moffat_results_F555W}
\end{figure}

\begin{figure}
\resizebox{\hsize}{!}{\includegraphics{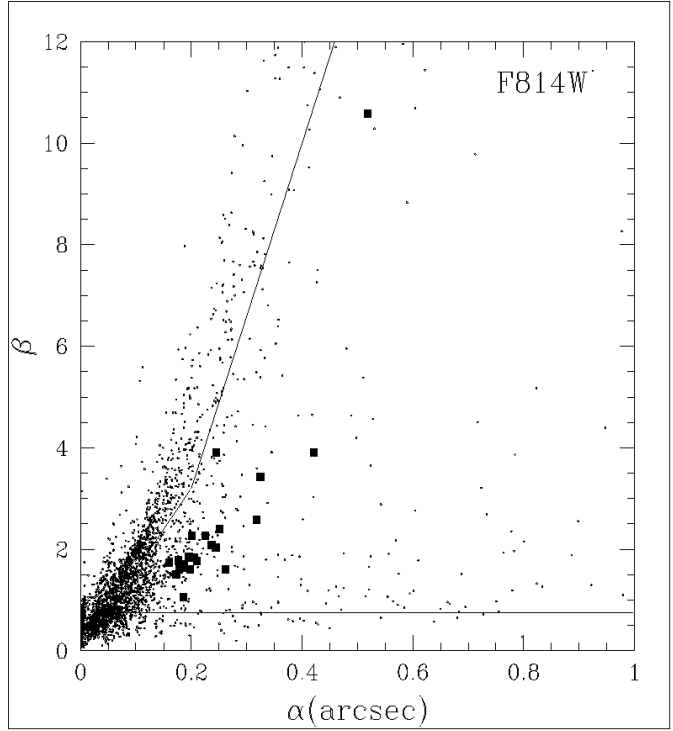}}
\caption{This figure shows the best-fitting Moffatian $\alpha$ and
$\beta$ parameters for each object (small circles) on the F814W
images.  Simulated data (see Fig.~\ref{FIGURE:moffat_art}) suggest
that objects that lie inside the wedge formed by the solid lines are
extended objects.  Therefore we consider any objects that lie inside
the wedge to to be GC candidates.  The solid squares show the
locations of the GC candidates from Table~\ref{TABLE:objects}.}
\label{FIGURE:moffat_results_F814W}
\end{figure}

	Figs.~\ref{FIGURE:moffat_results_F555W}
and~\ref{FIGURE:moffat_results_F814W} show the Moffatian $\alpha$ and
$\beta$ parameters for the $\sim$ 3800 objects that were successfully
fit by Moffatians.  There are 403 objects with Moffat parameters that
lie inside the wedge (see Fig.~\ref{FIGURE:moffat_art}).  Our
simulated data suggest that these are extended objects such as GCs,
background galaxies, dust features, open clusters, star forming
regions, or blended stars.

\subsubsection{Fitting Michie--King Models\label{SECTION:king_models}}

        We fit a two-dimensional, PSF-convolved, single-mass
Michie--King model to each of the 403 GC candidate using software
developed by Holland (\cite{H97}).  This software assumes that the
surface brightness profile along the effective radius axis of a GC
candidate with an ellipticity of $\epsilon$ and a position angle of
$\theta_0$ has a King profile with a concentration of $c$ and a core
radius of $r_c$.  It then builds a two-dimensional model based on this
surface brightness profile, $\epsilon$, and $\theta_0$.  The
two-dimensional model is convolved with the appropriate PSF for the
location on the CCD and a chi-square minimization is performed between
the PSF-convolved model and the original data image.  The software
uses CERN's {\sc minuit} function minimization package to fit
simultaneously the concentration, core radius, total flux in the
object, ellipticity, position angle, and mean background.  Objects
located within 32 pixels of the edge of a CCD ($= 3\farcs2$ for the
WFC and $1\farcs6$ for the PC) were not fit to avoid the edges of the
CCD biasing the fits.  Once a best fit had been determined, the King
tidal radius, $r_t$, and the half-mass radius, $r_h$, of the model
were computed.

        Separate fits were made to the F555W images and the F814W
images and an object was considered to be GC candidate only if a
Michie--King model could be fit in both colors.  We were able to fit
Michie--King models to 98 of the 403 potential GC candidates.  Mean
structural parameters were calculated for these object by taking the
mean of the values found in each filter.  Four objects (\#8, \#113,
\#128, and \#129) (see
Tables~\ref{TABLE:objects},~\ref{TABLE:iffy_objects1},~and~\ref{TABLE:iffy_objects2})
were identified on multiple fields.  In these cases we computed the
mean of the structural parameters measured in each field.

\begin{figure}
\resizebox{\hsize}{!}{\includegraphics{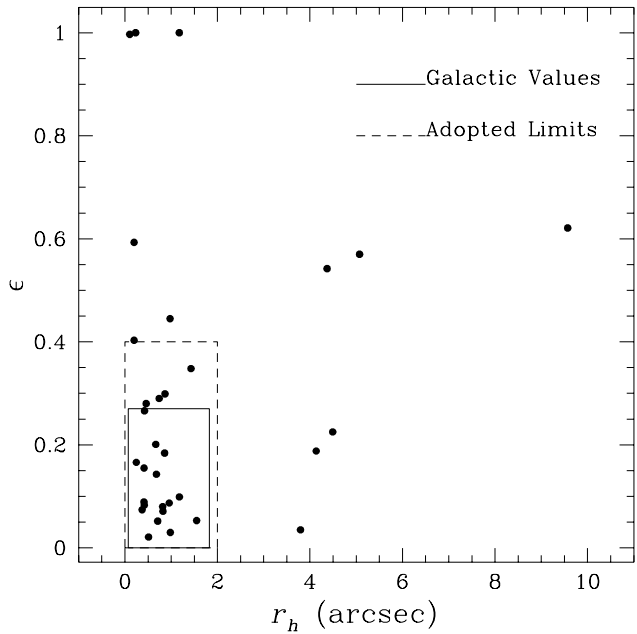}}
\caption{The ellipticity vs.\ half-mass radius of the best-fitting
single-mass Michie--King model for each object where a Michie--King
model was successfully fit.  Objects with $r_h > 10\arcsec$ ($\sim
175$ pc) have not been plotted. The solid box in the lower left of the
plot shows the region occupied by Galactic GCs.  Based on this plot we
have assumed that any object with $r_h < 2\arcsec$ ($\sim 35$ pc) and
$\epsilon < 0.4$ (the dashed box) is a GC candidate in the \object{NGC
5128} system.}
\label{FIGURE:rh_ell}
\end{figure}

	  We elected to separate GC candidates from background
galaxies based on their fitted ellipticities and half-mass radii (see
Fig.~\ref{FIGURE:rh_ell}).  Half-mass radii are preferred to tidal
radii or core radii because Fokker--Planck models of spherical stellar
systems show that half-mass radii remain reasonably constant over
periods of several Gyr (e.g.~Cohn~\cite{C79}; Takahashi~\cite{T97}),
making it a unique length scale for GCs.  The mass interior to the
half-mass radius tends to undergo a gravo-thermal collapse and become
concentrated at the center of the GC over time (i.e.~core-collapse),
which results in the core radius shrinking.  Meanwhile, the mass
exterior to the half-mass radius tends to expand outwards, causing the
tidal radius to grow.  Since we are interested in finding young,
intermediate-age, and old GCs in \object{NGC 5128}, it is useful to
have a selection criterion that does not depend on the age of the GC
candidate.  Galactic GCs have half-mass radii of approximately $1.3 <
r_h < 31.9$ pc (W. Harris~\cite{H96}), which corresponds to $0\farcs07
< r_h < 1\farcs83$ at the distance of \object{NGC 5128}.  There is no
evidence that the radius of a Galactic GC depends on its mass
(van~den~Bergh et~al.~\cite{VM91}).  Therefore, we have assumed that
only objects with $r_h \le 2\arcsec$ ($\sim 35$ pc at the distance of
\object{NGC 5128}) were GC candidates.  It is possible that some of
the objects in Fig.~\ref{FIGURE:rh_ell} that have high ellipticities
and low half-mass radii are double clusters.  However,
Innanen~et~al.~\cite{IH83} have shown that a binary GC could not
survive a single Galactic orbit in the \object{Milky Way} so it is
unlikely that there are any old, or intermediate-age double GCs in
\object{NGC 5128}.  It is possible that very young multiple GCs that
formed within the last $\sim 100-200$ Gyr could have survived to the
present day, but we are unable to differentiate between them and
background galaxies.

	The most elliptical Galactic GC is \object{M19} with $\epsilon
= 0.27$ (White \& Shawl~\cite{WS87}) and the most elliptical GC known
is \object{NGC 2193} in the \object{Large Magellanic Cloud} (LMC)
which has $\epsilon = 0.33$ (Geisler \& Hodge~\cite{GH80}).  Geisler
\& Hodge~(\cite{GH80}) modelled the distribution of observed
ellipticities for 25 GCs in the \object{LMC} and found that it was
unlikely that the largest true ellipticity exceeded $\epsilon = 0.4$.
The \object{LMC} contains both dynamically young and dynamically old
GCs, so the largest ellipticity seen in the LMC is a reasonable
estimate of the largest ellipticity that we can expect to see in
\object{NGC 5128}.  Therefore, only objects with $\epsilon \le 0.4$
were considered to be GC candidates.

	The final step was to examine visually the WFPC2 images of
each GC candidate to ensure that the Michie--King model fits looked
realistic.  We found that $\sim$ 20\% of the objects were either
located on diffraction spikes from saturated stars, or exhibited
unusually large residuals when the best-fitting Michie--King models
were subtracted.  These spurious identifications were discarded.

	 Fig.~\ref{FIGURE:rh_ell} shows the measured half-mass radii
and ellipticities for the surviving objects in \object{NGC 5128} and
Table~\ref{TABLE:objects} shows the final list of GC candidates that
we find in the central regions of NGC 5128.  The second and third
columns show the J2000 coordinates of the objects as determined using
the IRAF/STSDAS (v2.0.1) task {\sc stsdas.toolbox.imgtools.xy2rd}.
Column 4 is the observed (projected) distance of the GC candidate from
the center of \object{NGC 5128} in arcminutes.  The center of
\object{NGC 5128} was taken to be $\alpha_\mathrm{J2000} =
13^\mathrm{h} 25^\mathrm{m} 27\fs3$, $\delta_\mathrm{J2000} = -43\degr
01\arcmin 09\arcsec$ (Johnston et~al.~\cite{JF95}).  Columns 5 and 6
give the field (from Table~\ref{TABLE:obs_log}) and CCD that the
object was found on.  Columns 7 and 8 give the $X$ and $Y$ coordinates
(in pixels) on the CCD\@.  Column 9 lists the identification number of
the object in Table~1 of Minniti et~al.~(\cite{MA96}).


%
%

\begin{table*}
\begin{center}
\caption[]{The GC candidates in the central regions of \object{NGC 5128}.}
\smallskip
\begin{tabular}{rcccccccr}
\hline
\hline
\vspace{0.1 em}
ID & $\alpha_{J2000}$ & $\delta_{J2000}$ & $D$ & Field & CCD & $X$ & $Y$ & Other\\
\hline
\vspace{0.1 em}
  1 & $13^\mathrm{h}25^\mathrm{m}16\farcs1$ & $-42\degr59\arcmin45\arcsec$ & $2\farcm50$ & 3 & 3 & 703.990 & 570.660 & $\cdots$ \\
  2 & $13^\mathrm{h}25^\mathrm{m}16\farcs4$ & $-43\degr02\arcmin10\arcsec$ & $2\farcm26$ & 6 & 3 & 726.813 & 361.345 & $\cdots$  \\
  3 & $13^\mathrm{h}25^\mathrm{m}18\farcs1$ & $-43\degr01\arcmin59\arcsec$ & $1\farcm90$ & 6 & 3 & 516.154 & 406.027 & $\cdots$  \\
  4 & $13^\mathrm{h}25^\mathrm{m}19\farcs4$ & $-43\degr01\arcmin13\arcsec$ & $1\farcm47$ & 3 & 2 & 482.000 & 599.000 & $\cdots$  \\
  5 & $13^\mathrm{h}25^\mathrm{m}20\farcs5$ & $-43\degr00\arcmin52\arcsec$ & $1\farcm30$ & 3 & 2 & 310.789 & 429.262 & $\cdots$  \\
  6 & $13^\mathrm{h}25^\mathrm{m}25\farcs3$ & $-43\degr02\arcmin01\arcsec$ & $0\farcm95$ & 5 & 3 & 341.557 & 317.009 & $\cdots$  \\
  7 & $13^\mathrm{h}25^\mathrm{m}26\farcs6$ & $-42\degr59\arcmin27\arcsec$ & $1\farcm71$ & 1 & 3 & 686.555 & 752.859 & M27 \\
  8 & $13^\mathrm{h}25^\mathrm{m}27\farcs0$ & $-43\degr00\arcmin02\arcsec$ & $1\farcm12$ & 1 & 3 & 746.948 & 397.874 & M26  \\
  8 & $13^\mathrm{h}25^\mathrm{m}27\farcs0$ & $-43\degr00\arcmin02\arcsec$ & $1\farcm12$ & 2 & 3 & 171.527 & 228.000 & M26 \\
  9 & $13^\mathrm{h}25^\mathrm{m}27\farcs1$ & $-42\degr59\arcmin40\arcsec$ & $1\farcm49$ & 2 & 3 & 101.013 & 438.000 & M25 \\
 10 & $13^\mathrm{h}25^\mathrm{m}27\farcs2$ & $-42\degr59\arcmin28\arcsec$ & $1\farcm69$ & 1 & 3 & 621.989 & 725.544 & M24 \\
 11 & $13^\mathrm{h}25^\mathrm{m}27\farcs2$ & $-43\degr01\arcmin54\arcsec$ & $0\farcm75$ & 4 & 3 & 672.131 & 515.204 & $\cdots$  \\
 12 & $13^\mathrm{h}25^\mathrm{m}27\farcs8$ & $-42\degr59\arcmin23\arcsec$ & $1\farcm77$ & 1 & 3 & 551.791 & 755.891 & M23 \\
 13 & $13^\mathrm{h}25^\mathrm{m}28\farcs5$ & $-43\degr02\arcmin57\arcsec$ & $1\farcm81$ & 5 & 2 & 432.822 & 211.852 & $\cdots$  \\
 14 & $13^\mathrm{h}25^\mathrm{m}29\farcs2$ & $-43\degr00\arcmin16\arcsec$ & $0\farcm94$ & 1 & 3 & 553.540 & 198.363 & M20 \\
 15 & $13^\mathrm{h}25^\mathrm{m}29\farcs8$ & $-43\degr00\arcmin07\arcsec$ & $1\farcm12$ & 1 & 3 & 463.865 & 268.523 & M18 \\
 16 & $13^\mathrm{h}25^\mathrm{m}30\farcs2$ & $-42\degr59\arcmin36\arcsec$ & $1\farcm63$ & 1 & 3 & 337.483 & 548.311 & M17 \\
 17 & $13^\mathrm{h}25^\mathrm{m}31\farcs1$ & $-42\degr59\arcmin39\arcsec$ & $1\farcm65$ & 1 & 3 & 243.088 & 494.704 & $\cdots$  \\
 18 & $13^\mathrm{h}25^\mathrm{m}31\farcs5$ & $-43\degr00\arcmin04\arcsec$ & $1\farcm32$ & 1 & 3 & 280.111 & 236.550 & M15 \\
 19 & $13^\mathrm{h}25^\mathrm{m}31\farcs7$ & $-43\degr00\arcmin35\arcsec$ & $0\farcm97$ & 1 & 2 & 160.847 & 343.070 & $\cdots$  \\
 20 & $13^\mathrm{h}25^\mathrm{m}33\farcs7$ & $-43\degr01\arcmin20\arcsec$ & $1\farcm16$ & 1 & 2 & 655.839 & 265.242 & $\cdots$  \\
 21 & $13^\mathrm{h}25^\mathrm{m}34\farcs4$ & $-43\degr03\arcmin30\arcsec$ & $2\farcm67$ & 4 & 2 & 763.000 & 276.549 & $\cdots$  \\
\hline
\hline
\end{tabular}
\label{TABLE:objects}
\end{center}
\end{table*}

	Tables~\ref{TABLE:iffy_objects1}~and~\ref{TABLE:iffy_objects2}
lists the coordinates for the 61 extended objects with $r_h >
2\arcmin$ and $\epsilon > 0.4$.  Some of these objects may be GCs in
\object{NGC 5128} while others may be background galaxies with
structures similar to those of Michie--King models.  Six of these
objects have been previously identified as GCs by Minniti
et~al.~(\cite{MA96}) and Sharples~(\cite{S88}).


%
%

\begin{table*}
\begin{center}
\caption[]{Extended objects with $r_h > 2\arcsec$ and $\epsilon > 0.4$
in the central regions of \object{NGC 5128}}
\smallskip
\small
\begin{tabular}{rcccccccr}
\hline
\hline
\vspace{0.1 em}
ID & $\alpha_{J2000}$ & $\delta_{J2000}$ & $D$ & Field & CCD & $X$ & $Y$ & Other\\
\hline
\vspace{0.1 em}
 101 & $13^\mathrm{h}25^\mathrm{m}19\farcs3$ & $-43\degr02\arcmin30\arcsec$ & $2\farcm00$ & 6 & 3 &  508.484 &   64.127 &     G331 \\
 102 & $13^\mathrm{h}25^\mathrm{m}20\farcs0$ & $-43\degr01\arcmin56\arcsec$ & $1\farcm57$ & 6 & 3 &  307.668 &  352.171 & $\cdots$ \\
 103 & $13^\mathrm{h}25^\mathrm{m}20\farcs6$ & $-43\degr01\arcmin11\arcsec$ & $1\farcm25$ & 3 & 2 &  500.453 &  470.957 & $\cdots$ \\
 104 & $13^\mathrm{h}25^\mathrm{m}21\farcs0$ & $-42\degr59\arcmin25\arcsec$ & $2\farcm10$ & 3 & 3 &  128.163 &  607.910 & $\cdots$ \\
 105 & $13^\mathrm{h}25^\mathrm{m}21\farcs8$ & $-43\degr01\arcmin26\arcsec$ & $1\farcm07$ & 6 & 4 &  561.226 &   61.192 & $\cdots$ \\
 106 & $13^\mathrm{h}25^\mathrm{m}22\farcs1$ & $-43\degr00\arcmin39\arcsec$ & $1\farcm10$ & 3 & 2 &  234.000 &  214.221 & $\cdots$ \\
 107 & $13^\mathrm{h}25^\mathrm{m}22\farcs7$ & $-43\degr00\arcmin39\arcsec$ & $1\farcm00$ & 3 & 2 &  256.978 &  156.827 & $\cdots$ \\
 108 & $13^\mathrm{h}25^\mathrm{m}22\farcs8$ & $-43\degr00\arcmin40\arcsec$ & $0\farcm98$ & 2 & 2 &   99.928 &  720.877 & $\cdots$ \\
 109 & $13^\mathrm{h}25^\mathrm{m}23\farcs0$ & $-43\degr01\arcmin18\arcsec$ & $0\farcm82$ & 6 & 4 &  593.184 &  215.648 & $\cdots$ \\
 110 & $13^\mathrm{h}25^\mathrm{m}23\farcs1$ & $-43\degr00\arcmin50\arcsec$ & $0\farcm85$ & 3 & 2 &  377.582 &  144.000 & $\cdots$ \\
 111 & $13^\mathrm{h}25^\mathrm{m}23\farcs3$ & $-43\degr01\arcmin48\arcsec$ & $0\farcm99$ & 5 & 3 &  497.984 &  517.934 & $\cdots$ \\
 112 & $13^\mathrm{h}25^\mathrm{m}23\farcs6$ & $-43\degr01\arcmin34\arcsec$ & $0\farcm81$ & 5 & 3 &  421.420 &  632.818 & $\cdots$ \\
 113 & $13^\mathrm{h}25^\mathrm{m}23\farcs7$ & $-43\degr00\arcmin47\arcsec$ & $0\farcm78$ & 2 & 2 &  195.364 &  642.153 & $\cdots$ \\
 113 & $13^\mathrm{h}25^\mathrm{m}23\farcs7$ & $-43\degr00\arcmin47\arcsec$ & $0\farcm78$ & 3 & 2 &  369.000 &   69.211 & $\cdots$ \\
 114 & $13^\mathrm{h}25^\mathrm{m}24\farcs5$ & $-43\degr01\arcmin05\arcsec$ & $0\farcm54$ & 6 & 4 &  659.413 &  414.000 & $\cdots$ \\
 115 & $13^\mathrm{h}25^\mathrm{m}24\farcs9$ & $-43\degr00\arcmin20\arcsec$ & $0\farcm94$ & 2 & 3 &  444.626 &  122.783 & $\cdots$ \\
 116 & $13^\mathrm{h}25^\mathrm{m}25\farcs0$ & $-42\degr59\arcmin40\arcsec$ & $1\farcm55$ & 2 & 3 &  326.147 &  506.868 & $\cdots$ \\
 117 & $13^\mathrm{h}25^\mathrm{m}25\farcs3$ & $-43\degr02\arcmin01\arcsec$ & $0\farcm95$ & 6 & 4 &  101.986 &  296.981 & $\cdots$ \\
 118 & $13^\mathrm{h}25^\mathrm{m}25\farcs6$ & $-43\degr00\arcmin58\arcsec$ & $0\farcm38$ & 2 & 2 &  354.000 &  478.498 & $\cdots$ \\
 119 & $13^\mathrm{h}25^\mathrm{m}25\farcs9$ & $-43\degr00\arcmin57\arcsec$ & $0\farcm35$ & 2 & 2 &  358.270 &  438.501 & $\cdots$ \\
 120 & $13^\mathrm{h}25^\mathrm{m}26\farcs0$ & $-43\degr02\arcmin16\arcsec$ & $1\farcm14$ & 6 & 1 &  629.884 &  296.782 & $\cdots$ \\
 121 & $13^\mathrm{h}25^\mathrm{m}26\farcs5$ & $-42\degr59\arcmin55\arcsec$ & $1\farcm25$ & 2 & 3 &  207.559 &  311.875 & $\cdots$ \\
 122 & $13^\mathrm{h}25^\mathrm{m}26\farcs6$ & $-42\degr59\arcmin27\arcsec$ & $1\farcm71$ & 3 & 4 &  412.549 &  539.948 &      M27 \\
 123 & $13^\mathrm{h}25^\mathrm{m}26\farcs6$ & $-43\degr02\arcmin11\arcsec$ & $1\farcm04$ & 5 & 3 &  241.413 &  170.644 & $\cdots$ \\
 124 & $13^\mathrm{h}25^\mathrm{m}26\farcs8$ & $-43\degr00\arcmin12\arcsec$ & $0\farcm97$ & 2 & 3 &  224.417 &  144.083 & $\cdots$ \\
 125 & $13^\mathrm{h}25^\mathrm{m}26\farcs9$ & $-43\degr00\arcmin14\arcsec$ & $0\farcm93$ & 2 & 3 &  214.154 &  115.666 & $\cdots$ \\
 126 & $13^\mathrm{h}25^\mathrm{m}27\farcs1$ & $-43\degr00\arcmin06\arcsec$ & $1\farcm06$ & 1 & 3 &  744.088 &  355.292 & $\cdots$ \\
 127 & $13^\mathrm{h}25^\mathrm{m}27\farcs1$ & $-43\degr01\arcmin54\arcsec$ & $0\farcm75$ & 5 & 3 &  125.479 &  305.598 & $\cdots$ \\
 128 & $13^\mathrm{h}25^\mathrm{m}27\farcs3$ & $-42\degr59\arcmin28\arcsec$ & $1\farcm69$ & 2 & 3 &   45.380 &  554.534 &      M24 \\
 128 & $13^\mathrm{h}25^\mathrm{m}27\farcs3$ & $-42\degr59\arcmin28\arcsec$ & $1\farcm69$ & 3 & 4 &  386.949 &  604.233 &      M24 \\
 129 & $13^\mathrm{h}25^\mathrm{m}27\farcs3$ & $-43\degr01\arcmin54\arcsec$ & $0\farcm75$ & 5 & 3 &  113.000 &  300.561 & $\cdots$ \\
 129 & $13^\mathrm{h}25^\mathrm{m}27\farcs3$ & $-43\degr01\arcmin54\arcsec$ & $0\farcm75$ & 6 & 4 &   87.553 &  525.160 & $\cdots$ \\
 130 & $13^\mathrm{h}25^\mathrm{m}27\farcs4$ & $-43\degr01\arcmin28\arcsec$ & $0\farcm31$ & 2 & 2 &  711.369 &  371.000 & $\cdots$ \\
 131 & $13^\mathrm{h}25^\mathrm{m}27\farcs7$ & $-43\degr00\arcmin13\arcsec$ & $0\farcm93$ & 1 & 3 &  698.887 &  273.893 & $\cdots$ \\
 132 & $13^\mathrm{h}25^\mathrm{m}27\farcs8$ & $-42\degr59\arcmin23\arcsec$ & $1\farcm77$ & 2 & 4 &  582.795 &   98.323 & $\cdots$ \\
 133 & $13^\mathrm{h}25^\mathrm{m}28\farcs3$ & $-43\degr00\arcmin19\arcsec$ & $0\farcm85$ & 1 & 3 &  658.173 &  199.939 & $\cdots$ \\
 134 & $13^\mathrm{h}25^\mathrm{m}28\farcs3$ & $-43\degr01\arcmin38\arcsec$ & $0\farcm51$ & 5 & 4 &  408.000 &  133.199 & $\cdots$ \\
 135 & $13^\mathrm{h}25^\mathrm{m}28\farcs5$ & $-43\degr01\arcmin23\arcsec$ & $0\farcm30$ & 5 & 4 &  536.658 &  206.354 & $\cdots$ \\
 136 & $13^\mathrm{h}25^\mathrm{m}29\farcs0$ & $-43\degr00\arcmin43\arcsec$ & $0\farcm52$ & 2 & 2 &  323.925 &   78.964 & $\cdots$ \\
 137 & $13^\mathrm{h}25^\mathrm{m}29\farcs2$ & $-43\degr00\arcmin16\arcsec$ & $0\farcm94$ & 2 & 1 &  162.995 &   99.487 & $\cdots$ \\
 138 & $13^\mathrm{h}25^\mathrm{m}29\farcs2$ & $-43\degr00\arcmin22\arcsec$ & $0\farcm85$ & 1 & 3 &  570.063 &  142.000 & $\cdots$ \\
 139 & $13^\mathrm{h}25^\mathrm{m}29\farcs2$ & $-43\degr01\arcmin28\arcsec$ & $0\farcm45$ & 4 & 3 &  374.195 &  675.000 & $\cdots$ \\
 140 & $13^\mathrm{h}25^\mathrm{m}29\farcs2$ & $-43\degr02\arcmin29\arcsec$ & $1\farcm37$ & 4 & 3 &  599.716 &  108.340 & $\cdots$ \\
\hline
\hline
\end{tabular}
\label{TABLE:iffy_objects1}
\end{center}
\end{table*}


%
%

\begin{table*}
\begin{center}
\caption[]{Extended objects with $r_h > 2\arcsec$ and $\epsilon > 0.4$
in the central regions of \object{NGC 5128}, continued.}
\smallskip
\small
\begin{tabular}{rcccccccr}
\hline
\hline
\vspace{0.1 em}
ID & $\alpha_{J2000}$ & $\delta_{J2000}$ & $D$ & Field & CCD & $X$ & $Y$ & Other\\
\hline
\vspace{0.1 em}
 141 & $13^\mathrm{h}25^\mathrm{m}29\farcs2$ & $-43\degr00\arcmin38\arcsec$ & $0\farcm61$ & 1 & 2 &  109.346 &  606.980 & $\cdots$ \\
 142 & $13^\mathrm{h}25^\mathrm{m}29\farcs7$ & $-43\degr01\arcmin14\arcsec$ & $0\farcm42$ & 1 & 2 &  470.698 &  664.338 & $\cdots$ \\
 143 & $13^\mathrm{h}25^\mathrm{m}29\farcs8$ & $-43\degr00\arcmin07\arcsec$ & $1\farcm12$ & 2 & 4 &   95.982 &  190.565 &      M18 \\
 144 & $13^\mathrm{h}25^\mathrm{m}30\farcs0$ & $-43\degr02\arcmin12\arcsec$ & $1\farcm15$ & 4 & 3 &  460.000 &  234.000 & $\cdots$ \\
 145 & $13^\mathrm{h}25^\mathrm{m}30\farcs1$ & $-43\degr02\arcmin09\arcsec$ & $1\farcm11$ & 4 & 3 &  430.157 &  264.000 & $\cdots$ \\
 146 & $13^\mathrm{h}25^\mathrm{m}30\farcs3$ & $-43\degr01\arcmin12\arcsec$ & $0\farcm53$ & 1 & 2 &  466.932 &  598.975 & $\cdots$ \\
 147 & $13^\mathrm{h}25^\mathrm{m}30\farcs6$ & $-43\degr00\arcmin24\arcsec$ & $0\farcm95$ & 1 & 3 &  432.260 &   75.170 & $\cdots$ \\
 148 & $13^\mathrm{h}25^\mathrm{m}31\farcs1$ & $-43\degr01\arcmin19\arcsec$ & $0\farcm69$ & 5 & 4 &  476.273 &  487.763 & $\cdots$ \\
 149 & $13^\mathrm{h}25^\mathrm{m}31\farcs2$ & $-43\degr01\arcmin19\arcsec$ & $0\farcm71$ & 1 & 2 &  567.825 &  517.798 & $\cdots$ \\
 150 & $13^\mathrm{h}25^\mathrm{m}31\farcs3$ & $-43\degr00\arcmin37\arcsec$ & $0\farcm89$ & 1 & 2 &  169.525 &  384.000 & $\cdots$ \\
 151 & $13^\mathrm{h}25^\mathrm{m}31\farcs5$ & $-43\degr00\arcmin05\arcsec$ & $1\farcm30$ & 2 & 4 &   66.422 &  375.512 &      M15 \\
 152 & $13^\mathrm{h}25^\mathrm{m}31\farcs8$ & $-43\degr01\arcmin56\arcsec$ & $1\farcm12$ & 5 & 4 &  100.413 &  429.480 & $\cdots$ \\
 153 & $13^\mathrm{h}25^\mathrm{m}32\farcs6$ & $-43\degr01\arcmin27\arcsec$ & $0\farcm99$ & 5 & 4 &  345.530 &  611.654 & $\cdots$ \\
 154 & $13^\mathrm{h}25^\mathrm{m}33\farcs0$ & $-42\degr59\arcmin05\arcsec$ & $2\farcm31$ & 1 & 4 &  761.480 &  131.378 &      M12 \\
 155 & $13^\mathrm{h}25^\mathrm{m}33\farcs1$ & $-42\degr59\arcmin05\arcsec$ & $2\farcm32$ & 2 & 4 &  595.554 &  709.645 & $\cdots$ \\
 156 & $13^\mathrm{h}25^\mathrm{m}33\farcs2$ & $-43\degr01\arcmin37\arcsec$ & $1\farcm15$ & 5 & 4 &  224.435 &  632.127 & $\cdots$ \\
 157 & $13^\mathrm{h}25^\mathrm{m}33\farcs6$ & $-43\degr01\arcmin20\arcsec$ & $1\farcm14$ & 5 & 4 &  373.122 &  735.354 & $\cdots$ \\
 158 & $13^\mathrm{h}25^\mathrm{m}34\farcs6$ & $-43\degr02\arcmin37\arcsec$ & $1\farcm96$ & 4 & 2 &  280.000 &   74.000 & $\cdots$ \\
 159 & $13^\mathrm{h}25^\mathrm{m}35\farcs9$ & $-42\degr59\arcmin50\arcsec$ & $2\farcm04$ & 1 & 4 &  233.755 &  300.782 & $\cdots$ \\
 150 & $13^\mathrm{h}25^\mathrm{m}37\farcs0$ & $-43\degr01\arcmin29\arcsec$ & $1\farcm78$ & 4 & 4 &  360.651 &  495.008 & $\cdots$ \\
 161 & $13^\mathrm{h}25^\mathrm{m}37\farcs8$ & $-43\degr01\arcmin18\arcsec$ & $1\farcm90$ & 4 & 4 &  433.330 &  620.504 & $\cdots$ \\
\hline
\hline
\end{tabular}
\label{TABLE:iffy_objects2}
\end{center}
\end{table*}

	Figs.~\ref{FIGURE:f1_chart} through {\ref{FIGURE:f6_chart}}
show the locations of the 21 GC candidates on the F814W-band WFPC2
images.  Only objects that pass all of the criteria described above
are marked on these figures.  Objects (such as \#15) were only marked
on the fields that they were identified as GC candidates in.  In most
of the cases where a GC candidate is present in multiple fields, but
only identified in one field, the GC candidate was located very near
the edge of one of the CCDs.  Spatial variations in the PSF are
largest near the edges of the CCDs so the Michie-King model fits are
less reliable near the edges of the CCDs.

\begin{figure}
\caption{A finding chart for Field 1.  The GC candidates are circled
with their identification numbers (see Table~\ref{TABLE:objects})
printed near each object.}
\label{FIGURE:f1_chart}
\end{figure}

\begin{figure}
\caption{A finding chart for Field 2.}
\label{FIGURE:f2_chart}
\end{figure}

\begin{figure}
\caption{A finding chart for Field 3.}
\label{FIGURE:f3_chart}
\end{figure}

\begin{figure}
\caption{A finding chart for Field 4.}
\label{FIGURE:f4_chart}
\end{figure}

\begin{figure}
\caption{A finding chart for Field 5.}
\label{FIGURE:f5_chart}
\end{figure}

\begin{figure}
\caption{A finding chart for Field 6.}
\label{FIGURE:f6_chart}
\end{figure}

	The structural parameters of the best-fitting Michie--King
models, as well as the fitted ellipticities and position angles, for
each GC candidate are listed in Table~\ref{TABLE:structure}.  The
first column contains the object ID (from Table~\ref{TABLE:objects}).
The various radii are given in units of seconds of arc, and the
position angles are measured in degrees with $\theta_0 = 0\degr$ being
north and $\theta_0$ increasing to the east.  The $\chi^2_{\nu}$
values are the reduced goodness of fit values returned by the fitting
software.  The uncertainties ($\sigma$) are the standard deviations in
the values for the parameters that were measured in each filter.  The
position angles for GC candidates with small ellipticities ($< 0.05$)
are not reliable.  All of the $\chi^2_{\nu}$ are significantly less
than one, which suggests that the formal uncertainties in the model's
parameters are not reliable.  Therefore, we have elected to estimate
the uncertainties in the fits through monte-carlo simulations as
described in Sect.~\ref{SECTION:king_errors}.


%
%

\begin{table*}
\begin{center}
\caption{The best-fitting structural parameters for the NGC 5128 GC
candidates.  The values are the means of the structural parameters
derived from the F555W and F814W images.}
\smallskip
\begin{tabular}{rcccccccc}
\hline
\hline
\vspace{0.1 em}
ID & $W_0 \pm \sigma$ & $r_c \pm \sigma$ & $r_h \pm \sigma$ & $r_t \pm \sigma$ & $c \pm \sigma$ & $\epsilon \pm \sigma$ & $\theta_0 \pm \sigma$ & $\chi^2_{\nu}$ \\
\hline
\vspace{0.1 em}
  1 & $6.2 \pm \cdots$ & $0\farcs067 \pm 0\farcs001$ & $0\farcs681 \pm 0\farcs023$ & $1\farcs317 \pm 0\farcs045$ & $1.3 \pm 0.0$     & $0.143 \pm 0.015$ &  $-7\fdg9 \pm  2\fdg8$ & 0.134 \\
  2 & $6.8 \pm 0.3$    & $0\farcs053 \pm 0\farcs006$ & $0\farcs816 \pm 0\farcs080$ & $1\farcs582 \pm 0\farcs157$ & $1.5 \pm 0.1$     & $0.080 \pm 0.008$ & $+64\fdg2 \pm  3\fdg9$ & 0.171 \\
  3 & $4.1 \pm 1.1$    & $0\farcs113 \pm 0\farcs035$ & $0\farcs423 \pm 0\farcs057$ & $0\farcs801 \pm 0\farcs117$ & $0.9 \pm 0.2$     & $0.266 \pm 0.058$ & $+72\fdg4 \pm 19\fdg7$ & 0.116 \\
  4 & $8.6 \pm 0.2$    & $0\farcs024 \pm 0\farcs017$ & $1\farcs177 \pm 0\farcs726$ & $2\farcs286 \pm 1\farcs411$ & $2.0 \pm 0.1$     & $0.099 \pm 0.136$ &  $+4\fdg3 \pm 16\fdg4$ & 0.323 \\
  5 & $6.2 \pm 4.1$    & $0\farcs070 \pm 0\farcs076$ & $0\farcs741 \pm 0\farcs564$ & $1\farcs425 \pm 1\farcs107$ & $1.4 \pm 1.0$     & $0.290 \pm 0.132$ & $+71\fdg7 \pm  2\fdg5$ & 0.171 \\
  6 & $6.2 \pm 0.2$    & $0\farcs140 \pm 0\farcs001$ & $1\farcs428 \pm 0\farcs167$ & $2\farcs761 \pm 0\farcs326$ & $1.3 \pm 0.1$     & $0.348 \pm 0.123$ & $-28\fdg9 \pm  0\fdg7$ & 0.142 \\
  7 & $3.5 \pm 1.2$    & $0\farcs149 \pm 0\farcs101$ & $0\farcs413 \pm 0\farcs122$ & $0\farcs777 \pm 0\farcs219$ & $0.8 \pm 0.2$     & $0.155 \pm 0.023$ & $-44\fdg9 \pm 13\fdg0$ & 0.086 \\
  8 & $9.0 \pm 0.2$    & $0\farcs014 \pm 0\farcs006$ & $0\farcs452 \pm 0\farcs078$ & $0\farcs871 \pm 0\farcs163$ & $2.0 \pm 0.1$     & $0.140 \pm 0.084$ & $-30\fdg7 \pm 32\fdg9$ & 0.331 \\
  9 & $7.3 \pm 0.1$    & $0\farcs033 \pm 0\farcs001$ & $0\farcs708 \pm 0\farcs037$ & $1\farcs375 \pm 0\farcs071$ & $1.6 \pm 0.0$     & $0.052 \pm 0.059$ & $-79\fdg3 \pm 78\fdg5$ & 0.097 \\
 10 & $5.5 \pm 0.6$    & $0\farcs091 \pm 0\farcs015$ & $0\farcs667 \pm 0\farcs100$ & $1\farcs283 \pm 0\farcs201$ & $1.1 \pm 0.1$     & $0.201 \pm 0.004$ & $-54\fdg3 \pm  6\fdg2$ & 0.117 \\
 11 & $6.6 \pm 1.0$    & $0\farcs102 \pm 0\farcs001$ & $1\farcs551 \pm 0\farcs920$ & $3\farcs008 \pm 1\farcs797$ & $1.4 \pm 0.3$     & $0.053 \pm 0.004$ & $+21\fdg5 \pm 16\fdg0$ & 0.207 \\
 12 & $4.2 \pm 1.0$    & $0\farcs092 \pm 0\farcs023$ & $0\farcs371 \pm 0\farcs062$ & $0\farcs704 \pm 0\farcs126$ & $0.9 \pm 0.2$     & $0.074 \pm 0.002$ & $+80\fdg2 \pm  \cdots$ & 0.082 \\
 13 & $7.3 \pm 0.1$    & $0\farcs039 \pm 0\farcs001$ & $0\farcs822 \pm 0\farcs029$ & $1\farcs597 \pm 0\farcs057$ & $1.6 \pm 0.0$     & $0.071 \pm 0.023$ & $-47\fdg9 \pm 78\fdg8$ & 0.066 \\
 14 & $8.0 \pm 0.8$    & $0\farcs015 \pm 0\farcs009$ & $0\farcs460 \pm 0\farcs045$ & $0\farcs892 \pm 0\farcs087$ & $1.8 \pm 0.2$     & $0.280 \pm 0.187$ & $+74\fdg3 \pm 11\fdg5$ & 0.361 \\
 15 & $4.8 \pm 0.1$    & $0\farcs084 \pm 0\farcs006$ & $0\farcs419 \pm 0\farcs014$ & $0\farcs799 \pm 0\farcs026$ & $1.0 \pm 0.0$     & $0.083 \pm 0.000$ & $-13\fdg3 \pm  2\fdg7$ & 0.120 \\
 16 & $7.2 \pm 0.4$    & $0\farcs044 \pm 0\farcs009$ & $0\farcs867 \pm 0\farcs041$ & $1\farcs684 \pm 0\farcs081$ & $1.6 \pm 0.1$     & $0.299 \pm 0.033$ & $-54\fdg1 \pm  2\fdg9$ & 0.080 \\
 17 & $4.7 \pm 1.6$    & $0\farcs048 \pm 0\farcs018$ & $0\farcs245 \pm 0\farcs076$ & $0\farcs468 \pm 0\farcs154$ & $1.0 \pm 0.3$     & $0.166 \pm 0.094$ & $-32\fdg5 \pm  \cdots$ & 0.079 \\
 18 & $6.9 \pm 0.0$    & $0\farcs062 \pm 0\farcs000$ & $0\farcs982 \pm 0\farcs018$ & $1\farcs904 \pm 0\farcs034$ & $1.5 \pm 0.0$     & $0.030 \pm 0.016$ & $-50\fdg2 \pm 15\fdg6$ & 0.107 \\
 19 & $5.0 \pm \cdots$ & $0\farcs123 \pm 0\farcs010$ & $0\farcs711 \pm \cdots$     & $1\farcs360 \pm \cdots$     & $1.0 \pm \cdots$  & $0.292 \pm 0.015$ & $-10\fdg6 \pm 19\fdg7$ & 0.250 \\
 20 & $7.2 \pm 1.0$    & $0\farcs050 \pm 0\farcs025$ & $0\farcs958 \pm 0\farcs132$ & $1\farcs860 \pm 0\farcs260$ & $1.6 \pm 0.3$     & $0.087 \pm 0.063$ & $-26\fdg2 \pm  7\fdg1$ & 0.146 \\
 21 & $3.9 \pm 2.0$    & $0\farcs138 \pm 0\farcs058$ & $0\farcs511 \pm 0\farcs191$ & $0\farcs970 \pm 0\farcs382$ & $0.8 \pm 0.4$     & $0.021 \pm 0.029$ & $-51\fdg4 \pm 43\fdg6$ & 0.085 \\
\hline
\hline
\end{tabular}
\label{TABLE:structure}
\end{center}
\end{table*}

\subsection{Contamination\label{SECTION:contamination}}

	Our data will contain images of Galactic foreground stars, and
supergiants in \object{NGC 5128}.  From the work of Bahcall \&
Soneira~(\cite{BS81}) we expect to find $\sim 700$ Galactic stars with
$I \le 20$ in our fields.  The brightest stars in the halo of
\object{NGC 5128} have $V \sim 24.5$ (Soria et~al.~\cite{SM96}) while
the brightest young stars in the central regions of \object{NGC 5128}
are expected to have approximately $18 < V < 21$.  The morphological
selection criteria that we applied to our data are very effective at
rejecting stars (see
Figs.~\ref{FIGURE:moffat_art},~\ref{FIGURE:moffat_results_F555W},~and~\ref{FIGURE:moffat_results_F814W},),
and the \emph{HST} images show all of the 21 GC candidates to be
extended objects, so we believe that stellar contamination is not a
problem in our data.

	The galaxy counts of Tyson~(\cite{T88}) suggest that there
will be $\sim 10$ background galaxies in our images down to $I \sim
20$.  However, many of these galaxies will be obscured by the dust
lane, so we will detect significantly fewer than this.  The
morphological criteria that we applied to obtain our list of GC
candidates will reject any galaxies that are not morphologically
similar to the GCs found in the \object{Milky Way} or \object{LMC}.
AM97 used a comparison field located $30\arcmin$ northeast of the
nucleus to estimate that $\sim 20$\% of the objects that they detect
in their search for GCs in the inner regions of \object{NGC 5128} are
foreground stars or background galaxies.  Since the morphological
criteria that we applied are stricter than those of AM97, we believe
that 20\% is a reasonable upper limit on the amount of contamination
in our list of GC candidates.

\subsection{Uncertainties in the Structural Parameters\label{SECTION:king_errors}}

        The {\sc minuit} package provides an estimate of the formal
uncertainty in each parameter based on the covariance matrix of the
fit.  In general the formal uncertainties were $\sim 10$\% of the
best-fit value of each parameter.  This is consistent with the
uncertainties quoted in Table~\ref{TABLE:structure} that were
determined from the differences between the best fitting structural
parameters determined from the F555W images and the F814W images.

        In order to test the formal uncertainty estimates, and to look
for systematic differences between the recovered structural parameters
and the true structural parameters of the \object{NGC 5128} GC
candidates, we constructed a series of artificial GCs and added them
to the \object{NGC 5128} images.  The total of 810 artificial GCs were
added to the WFPC2 images with randomly assigned concentrations
between $0.67 \le c \le 2.74$, core radii between $0\farcs07 \le r_c
\le 0\farcs4$, ellipticities of $\epsilon = 0$, and magnitudes of $16
\le \mathrm{F555W} \le 22$.  We found that the recovered
concentrations for the brightest artificial GCs ($16 \le
\mathrm{F555W} \le 18$) were within $\sim 5$\% of their true values
95\% of the time.  For the faintest artificial GCs ($20 \le
\mathrm{F555W} \le 22$) the recovered values were within 22\% of the
true values 95\% of the time.  Systematic shifts were negligible for
artificial GCs with concentrations greater than $c \sim 0.9$, but grew
rapidly for artificial GCs with concentrations less than this.  The
uncertainties in the structural parameters increased as the
concentration decreased.  Other structural parameters behaved in very
similar manners.

       A second source of uncertainty in the fitted structural
parameters is the uncertainty in the PSF\@.  Each GC candidate was
fitted with a Michie--King model that was convolved by an estimate of
the PSF at the location of the object on the CCD\@.  We used PSFs that
were created by Peter Stetson (1996, private communication) from WFPC2
observations of stars in the Galactic GC \object{$\omega$ Centauri}.
However, the \object{NGC 5128} images were taken approximately three
years after the \object{$\omega$ Cen} ones, so long-term variations in
the focus of the WFPC2 (e.g.~Suchkov \& Casertano~\cite{SC97}) raise
the possibility of a mismatch between the actual and adopted PSFs. If
this is the case then the derived structural parameters would be in
error, as the fitted Michie--King models were convolved with a PSF
constructed using a slightly different focus from that of our images.
In order to estimate the possible errors introduced by uncertainties
in the PSF, we repeated the Michie--King model fitting procedure with
the \emph{wrong} PSFs.  In other words, we fit Michie--King models to
the GC candidates on the WF3 CCD using both the F555W and the F814W
PSFs from the WF2, WF3, and WF4 CCDs.  This gave us six estimates of
the structural parameters of each object obtained using six variations
of the WF PSF\@.  For each of these estimates we computed the
difference between $W_0$ as determined using the correct PSF, and the
five $W_0$s determined using the incorrect PSFs ($\Delta W_0$).  We
then computed the mean ($\overline{\Delta W_0}$) and standard error in
the mean of these five values for each GC candidate.  This gave us an
estimate of the systematic uncertainty in the value of $W_0$ that we
derived for each GC candidate.  Finally, we computed the mean,
standard error in the mean, and median of the individual
$\overline{\Delta W_0}$ values for all the GC candidate.  This gave an
estimate of the systematic uncertainty for a typical GC candidate in
our data.  These values, along with analogous estimates of the
systematic errors in the other structural parameters, are presented in
Table~\ref{TABLE:PSF_errs}.  This technique is not mathematically
rigorous since the variations in the PSF from one CCD to another, and
from one filter to another, are not the same as the variations due to
changes in the focus of the \emph{HST} over a period of two or three
years.  However, Suchkov \& Casertano~(\cite{SC97}) find that
long-term changes in focus introduce changes of a few percent in the
photometric calibrations, which corresponds to changes of a few
percent in the shape of the PSF\@.  The variations in the PSF from one
CCD to the next can be much larger than this so we believe that the
systematic errors derived here represent an over-estimate of the true
systematic errors introduced by possible long-term variations in the
PSF\@.


%
%

\begin{table}
\begin{center}
\caption{Systematic errors due to uncertainties in the PSF.}
\smallskip
\begin{tabular}{llll}
\hline
\hline
\vspace{0.1 em}
Quantity & mean & se & median \\
\hline
\vspace{0.1 em}
$\overline{\Delta W_0}$      &     0.4      &     0.1       &     0.2 \\
$\overline{\Delta r_c}$      & $0\farcs010$ & $0\farcs003$ & $0\farcs004$ \\
$\overline{\Delta r_h}$      & $0\farcs079$ & $0\farcs026$ & $0\farcs030$ \\
$\overline{\Delta r_t}$      & $0\farcs154$ & $0\farcs051$ & $0\farcs060$ \\
$\overline{\Delta c}$        &     0.1      &     0.0      &     0.0 \\
$\overline{\Delta \epsilon}$ &    0.024     &    0.004     &    0.022 \\
$\overline{\Delta \theta_0}$ &  $31\fdg8$   &  $15\fdg2$   &  $7\fdg4$ \\
\hline
\hline
\end{tabular}
\label{TABLE:PSF_errs}
\end{center}
\end{table}


\section{Structural Parameters\label{SECTION:structure}}

\subsection{Core Radii\label{SECTION:core_radii}}

\begin{figure}
\resizebox{\hsize}{!}{\includegraphics{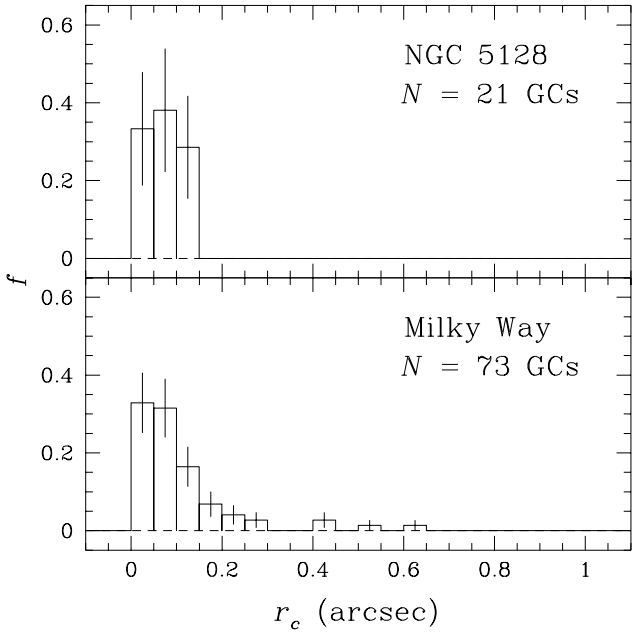}}
\caption{This figure compares the distribution of core radii for GC
candidates in \object{NGC 5128} with the distribution of core radii
for selected GCs (see Sect.~\ref{SECTION:core_radii}) in the
\object{Milky Way}.  The vertical axis is the fraction of the total
number of GCs and the error bars show the Poisson uncertainties in
each bin.}
\label{FIGURE:core_radii}
\end{figure}

        Fig.~\ref{FIGURE:core_radii} shows the distribution of King
core radii for the \object{NGC 5128} GC candidates and for the
\object{Milky Way} GCs.  The \object{Milky Way} GCs have been shifted
to the distance of \object{NGC 5128} and their core radii have been
converted to arcseconds ($1\arcsec = 17.45 \pm 0.97$ pc for $d = 3.6
\pm 0.2$ Mpc).  Galactic GCs with a central brightness cusp (which is
believed to be a signature of a collapsed core) have been omitted from
the sample.  The faintest GC candidate in our sample has an absolute
total magnitude of $M_{V,\mathrm{tot}} = -6.4 \pm 0.2$.  In order to
ensure that we are comparing similar objects we have excluded all
\object{Milky Way} GCs fainter than $M_{V,\mathrm{tot}} = -6.4$.

	In order to facilitate a comparison between the two GC systems
we plotted the fraction, $f_i = n_i / n$, of the total number of GCs
in each bin, $n_i$, where $n$ is the total number of GCs in each data
set.  The uncertainty in $f_i$ was computed from the Poisson
uncertainties in the number of GCs in each bin, and in the total
number of GCs using

\begin{equation}
\sigma_{f,i} = {1 \over n} \sqrt{n_i + {n_i^2 \over  n}}.
\label{EQUATION:uncertainty}
\end{equation}

\noindent
The mean core radius for the 21 \object{NGC 5128} GC candidates is
$\overline{r_c} = 0\farcs07 \pm 0\farcs01$ (standard error) while the
mean for the 73 selected \object{Milky Way} GCs is $\overline{r_c} =
0\farcs11 \pm 0\farcs01$ (se).  A Kolmogorov--Smirnov (KS) test shows
that we can reject the hypothesis that the two samples are drawn from
the same distribution at the 46\% confidence level.  Therefore, there
is no evidence that the core radii of the GC candidates in \object{NGC
5128} are distributed differently from the core radii of the
\object{Milky Way} GCs.

	The most noticeable difference between the core radii of the
\object{NGC 5128} GC candidates and the core radii of the
\object{Milky Way} GCs in Fig.~\ref{FIGURE:core_radii} is the lack of
a tail extending to large core radii in the \object{NGC 5128} data.
This may be an artifact of the small number (21) of \object{NGC 5128}
GC candidates in our sample.  The mean of a distribution is sensitive
to the presence of tails and outliers, but the median is much more
robust against outliers. Therefore we computed the median core radius
for each data set. Both the \object{NGC 5128} GC candidates and the
\object{Milky Way} GCs have median core radii of $[r_c] = 0\farcs07$
($= 1.22$ pc, or $\sim 0.7$ pixels on the WF CCDs and 1.4 pixels on
the PC CCD).  The similarity in the median core radii suggests that
the ``typical'' core radius of a GC candidate in \object{NGC 5128} is
similar to that of the \object{Milky Way} GCs.

	There may be systematic biases in the core radii that we have
derived for the GC candidates in \object{NGC 5128}.  Fitting
Michie--King models to objects with core radii that are similar to the
pixel scales of the images requires that the centers of the objects be
accurately known.  Small errors in determining the center of a
candidate GC, and small systematic errors introduced by integrating
Michie--King model profiles over the area of a pixel, may be
sufficient to bias the fitted core radii towards smaller values.  In
addition to pixelation effects, the similarity between the core radii
and the FWHMs of the PSFs may also be biasing our fits toward smaller
core radii.

\subsection{Tidal Radii\label{SECTION:tidal_radii}}

	The tidal radius of a GC is affected by the gravitational
potential of its parent galaxy (e.g.,~Innanen et~al.~\cite{IH83};
Heggie \& Ramamani~\cite{HR95}).  In order to compare the tidal radii
of GC candidates in \object{NGC 5128} with those of GCs in the
\object{Milky Way} it is necessary to correct for the tidal fields of
both galaxies.  The first step is to normalize the tidal radius of
each GC candidate by its mass, ${\cal M}_\mathrm{cl}$, to get
$r_t/{\cal M}_\mathrm{cl}^{1/3}$.  If we assume that the gravitational
potentials of the \object{Milky Way} and \object{NGC 5128} can be
approximated by a spherical logarithmic potential of the form

\begin{equation}
\Phi = V_\mathrm{rot}^2 \ln(R^2 + R_s^2) + \ln(C),
\label{EQUATION:log_pot}
\end{equation}

\noindent
where $R$ is the galactocentric distance, $R_s$ is a scale length, and
$C$ is a constant, then we can compare the mean value of the
normalized tidal radii of the GC candidates using

\begin{equation}
{r_t \over {\cal M}_\mathrm{cl}^{1/3} } =
        {\left( {R_p \over V_\mathrm{rot}} \right)}^{2/3}
        {\left( {G \over 2 g(e)} \right)}^{1/3},
\label{EQUATION:tidal_radius}
\end{equation}

\noindent
where $V_\mathrm{rot}$ is the amplitude of the flat part of the
rotation curve of the galaxy, $G$ is the Newtonian gravitational
constant, and $g(e)$ is a slowly varying function of the orbital
eccentricity of the GC that has values of $g(0) = 1$ for circular
orbits.  The mean value of the normalized tidal radius of the GC
candidates in \object{NGC 5128}, $\langle r_t/{\cal
M}_\mathrm{cl}^{1/3} \rangle$, can then be related to the mean
normalized tidal radius of the \object{Milky Way} GCs by

\begin{eqnarray}
\nonumber
\left\langle{r_t \over {\cal M}_\mathrm{cl}^{1/3}}\right\rangle = &
  {\left({V_{\mathrm{rot},\mathrm{MW}} \over V_\mathrm{rot}}\right)}^{2/3}
  {\left({g_\mathrm{MW}(e) \over g(e)}\right)}^{1/3} 
  {\langle R_p^{2/3} \rangle \over \langle R_{p,\mathrm{MW}}^{2/3} \rangle} \\
 &       \left\langle{r_t \over 
               {\cal M}_\mathrm{cl}^{1/3}}\right\rangle_{\mathrm{MW}},
\label{EQUATION:ratio}
\end{eqnarray}

\noindent
where the subscript MW denotes the value for the \object{Milky Way}.

	Eq.~\ref{EQUATION:ratio} assumes that the shape of the
galactic potential is the same in both galaxies, but allows the total
mass, as parameterized by $V_\mathrm{rot}$, of each galaxy to vary.
It also requires a knowledge of the distribution of GC orbits in each
galaxy, as parameterized by $g(e)$ and $R_p$.  We assumed a rotation
velocity of $V_\mathrm{rot,MW} = 220\,\mathrm{km}\,\mathrm{s}^{-1}$
for the \object{Milky Way} and $V_\mathrm{rot} =
245\,\mathrm{km}\,\mathrm{s}^{-1}$ for \object{NGC 5128} (Hui
et~al.~\cite{HF95}).  The only information available on the
distribution of orbits for GC candidates in \object{NGC 5128} is the
projected radial distances of the GC candidates from the center of
\object{NGC 5128}, so we have assumed that the \object{NGC 5128} GC
system is dynamically similar to the \object{Milky Way} GC system.
This involves two assumptions about the nature of the GC orbits.
First, we assume that the mean eccentricity of the \object{NGC 5128}
GC orbits is the same as that for the \object{Milky Way} GCs
($\overline{e_\mathrm{MW}} = 0.6 \pm 0.1$, Odenkirchen
et~al.~\cite{OB97}).  Second, we assume that each GC is near the
apogalacticon of its orbit, $R_a$, so $R \sim R_a$.  This is a
reasonable assumption since a GC having $\epsilon \sim 0.6$ spends
$\sim 80$\% of its time in the outer half of its orbit.  An additional
complication is that the observed galactocentric distance for a
\object{NGC 5128} GC candidate is the projection of the true
galactocentric distance onto the plane of the sky.  If we assume that
the orbits are oriented randomly then $R_a$ for a particular object's
orbit is most likely to be observed to lie at a projected distance of
$D = (8/\pi^2) R_a$.  Therefore, we estimated the perigalactic
distances for each of the \object{NGC 5128} GC candidates from the
observed distance from the center of \object{NGC 5128} using

\begin{equation}
R_p = {1-e \over 1+e} {\pi^2 \over 8} D.
\label{EQUATION:perigalacticon}
\end{equation}

\begin{figure}
\resizebox{\hsize}{!}{\includegraphics{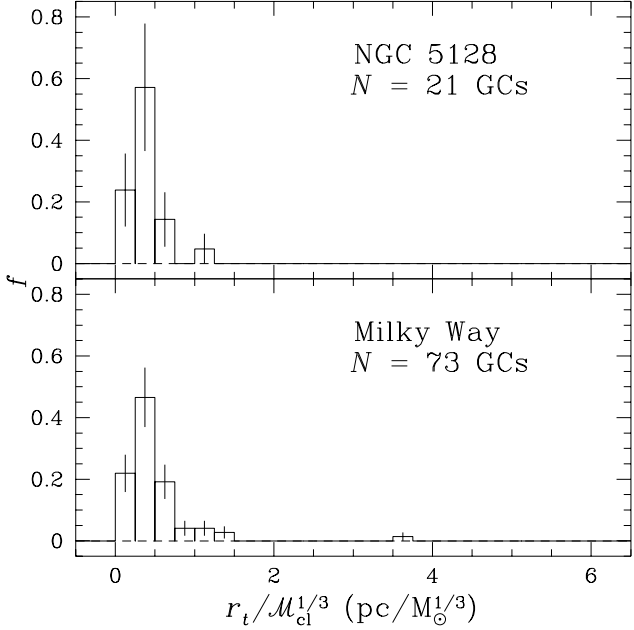}}
\caption{This figure compares the distribution of normalized tidal
radii (see Sect.~\ref{SECTION:tidal_radii}) for GC candidates in
\object{NGC 5128} with the distribution of normalized tidal radii
(after correcting for the difference in mass between the two galaxies)
for GCs in the \object{Milky Way}.  The vertical axis is the fraction
of the total number of GCs and the error bars show the Poisson
uncertainties in each bin.}
\label{FIGURE:tidal_radii}
\end{figure}

	Fig.~\ref{FIGURE:tidal_radii} shows the distribution of the
normalized tidal radii for the \object{NGC 5128} GC candidates and the
\object{Milky Way} GCs.  The cluster masses were computed from their
total $V$-band luminosities assuming a mass-to-light ratio of two.

	Our sample of GC candidates has $\langle r_t / {\cal
M}_\mathrm{cl}^{1/3} \rangle = 0.36 \pm 0.05$ (se)
$\mathrm{pc}/\mathrm{M}_{\sun}^{1/3}$ ($N=21$).  The mean normalized
tidal radius for 73 selected \object{Milky Way} GCs is $\langle r_t /
{\cal M}_\mathrm{cl}^{1/3} \rangle_\mathrm{MW} = 0.69 \pm 0.08$ (se)
$\mathrm{pc}/\mathrm{M}_{\sun}^{1/3}$.  The multiplicative factor in
Eq.~\ref{EQUATION:ratio} is 0.703, which yields a corrected $\langle
r_t / {\cal M}_\mathrm{cl}^{1/3} \rangle_\mathrm{MW}$ of $0.49 \pm
0.06$ (se) pc/ M$_{\sun}^{1/3}$.  A KS test says that we can reject
the hypothesis that the two samples are drawn from the same
distribution at the 74\% confidence level.  Therefore, there is
insufficient evidence to state that the distribution of the tidal
radii of the \object{NGC 5128} GC candidates differs from that of the
Galactic GCs \emph{if the difference in the masses of the two galaxies
is taken into account}.  However, we wish to stress that this
calculation assumes that the distribution of GC orbits are
statistically similar for both galaxies.

\subsection{Half-Mass Radii\label{SECTION:half_mass_radii}}

\begin{figure}
\resizebox{\hsize}{!}{\includegraphics{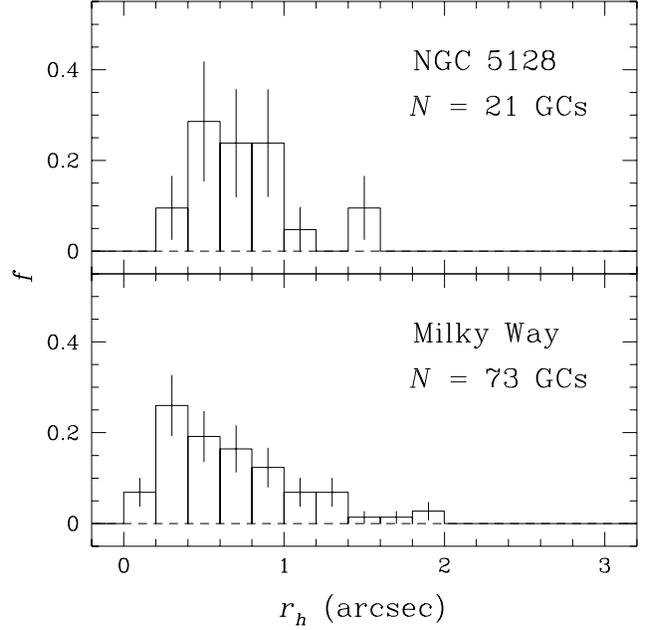}}
\caption{This figure compares the distribution of half-mass radii for
GC candidates in \object{NGC 5128} with the distribution of half-mass
radii for a subset of GCs in the \object{Milky Way}.  The vertical
axis is the fraction of the total number of GCs and the error bars
show the Poisson uncertainties in each bin.}
\label{FIGURE:half_mass_radii}
\end{figure}

        The distribution of half-mass radii for the \object{NGC 5128}
GC candidates is shown in Fig.~\ref{FIGURE:half_mass_radii} along with
the same distribution for our subsample of 73 \object{Milky Way} GCs.
The half-mass radii for the \object{Milky Way} GCs were determined by
computing a Michie--King model (with concentrations and core radii
taken from W. Harris~\cite{H96}) for each \object{Milky Way} GC\@.
This allowed us to make a direct comparison between the half-mass
radii of the best-fitting single-mass Michie--King models for the
\object{NGC 5128} GC candidates and the half-mass radii of the
best-fitting single-mass Michie--King models for the \object{Milky
Way} GCs.

	The mean half-mass radius for the 21 \object{NGC 5128} GC
candidates is $\overline{r_h} = 0\farcs73 \pm 0\farcs08$ (se) while
the mean for 73 selected \object{Milky Way} GCs is $\overline{r_h} =
0\farcs67 \pm 0\farcs05$ (se).  However, a KS test says that we can
reject the hypothesis that the two samples are drawn from the same
distribution at only the 74\% confidence level, so there is no
evidence that the distribution of half-mass radii in our sample of
\object{NGC 5128} GC candidates is different from that of the GCs in
the \object{Milky Way}.

\subsection{Ellipticities\label{SECTION:ellipticities}}

\begin{figure}
\resizebox{\hsize}{!}{\includegraphics{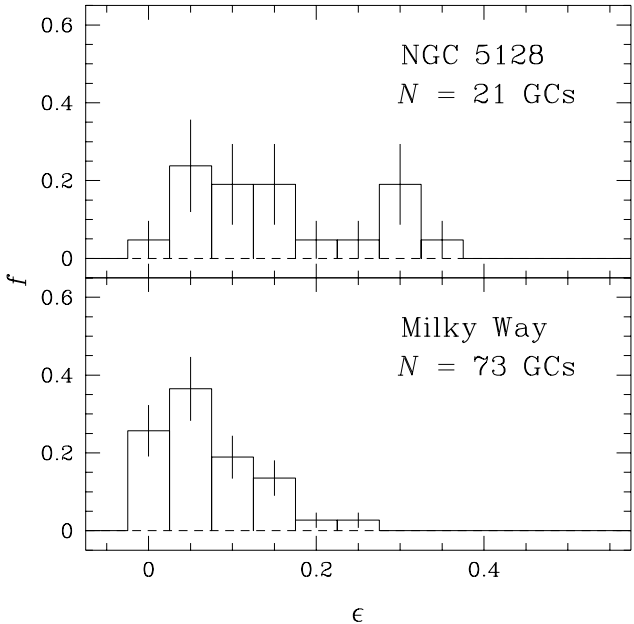}}
\caption{The upper panel shows the distribution of ellipticities for
the GC candidates in \object{NGC 5128}.  The lower panel shows the
distribution of ellipticities for the \object{Milky Way}'s GCs (from
White \& Shawl~\protect\cite{WS87}).}
\label{FIGURE:ell}
\end{figure}

        The distribution of ellipticities for the \object{NGC 5128} GC
candidates is shown in Fig.~\ref{FIGURE:ell}.  The 21 \object{NGC
5128} objects have $\overline{\epsilon} = 0.15 \pm 0.02$ (se) while
the White \& Shawl~(\cite{WS87}) sample of \object{Milky Way} GCs
yields $\overline{\epsilon} = 0.07 \pm 0.01$ (se) for the 73
\object{Milky Way} GCs.  A KS test says that we can reject the
hypothesis that the two samples are drawn from the same distribution
at the 99.7\% confidence level.  Therefore, we conclude that the
\object{NGC 5128} GC candidates may have a different distribution of
ellipticities from the \object{Milky Way} GCs.

	The \object{NGC 5128} GC candidates appear to be
systematically more elliptical than the \object{Milky Way} GCs.  There
appears to be a lack of objects with low ellipticities and an excess
of GC candidates with $\epsilon \sim 0.3$.  The lack of GC candidates
with $\epsilon \le 0.05$ is probably due to the elliptical PSF not
being fully removed from the data.  Another possible source of
ellipticity is the stochastic distribution of bright stars near the
center of the object.  Geisler \& Hodge~(\cite{GH80}) found that the
random placement of stars with respect to the adopted center of a GC
can introduce a systematic error in the observed ellipticity of
$+0.045 \pm 0.015$ for GCs which are intrinsically spherical.  This
effect acts to make nearly spherical GCs appear to be more elliptical
than they actually are.  They also found that this systematic error
decreases as the intrinsic ellipticity of the GCs increases.  This
would explain the lack of nearly circular GC candidates in \object{NGC
5128}, relative to the \object{Milky Way}.

        AM97 identified 125 GC candidates in the inner $2\farcm8
\times 2\farcm8$ of \object{NGC 5128}.  Table~\ref{TABLE:compare}
lists the ellipticities from those GC candidates in common between the
two studies.  The mean difference between our ellipticities and the
AM97 values is $\Delta \epsilon = \epsilon(\mathrm{this~work}) -
\epsilon(\mathrm{AM97}) = -0.024 \pm 0.041$.


%
%

\begin{table}
\begin{center}
\caption{A comparison of our ellipticities with those of AM97.}
\smallskip
\begin{tabular}{rccc}
\hline
\hline
\vspace{0.1 em}
ID & AM97 & Our $\epsilon \pm \sigma$ & AM97 $\epsilon$ \\
\hline
\vspace{0.1 em}
  9 & M25 & $0.052 \pm 0.059$ & 0.02 \\
 10 & M24 & $0.201 \pm 0.004$ & 0.09 \\
 12 & M23 & $0.074 \pm 0.002$ & 0.04 \\
 14 & M20 & $0.280 \pm 0.187$ & 0.06 \\
 15 & M18 & $0.083 \pm 0.000$ & 0.05 \\
 16 & M17 & $0.299 \pm 0.033$ & 0.14 \\
 18 & M15 & $0.030 \pm 0.016$ & 0.09 \\
\hline
\hline
\end{tabular}
\label{TABLE:compare}
\end{center}
\end{table}

\subsection{Correlations of Cluster Properties with Distance\label{correlations}}

\begin{figure}
\resizebox{\hsize}{!}{\includegraphics{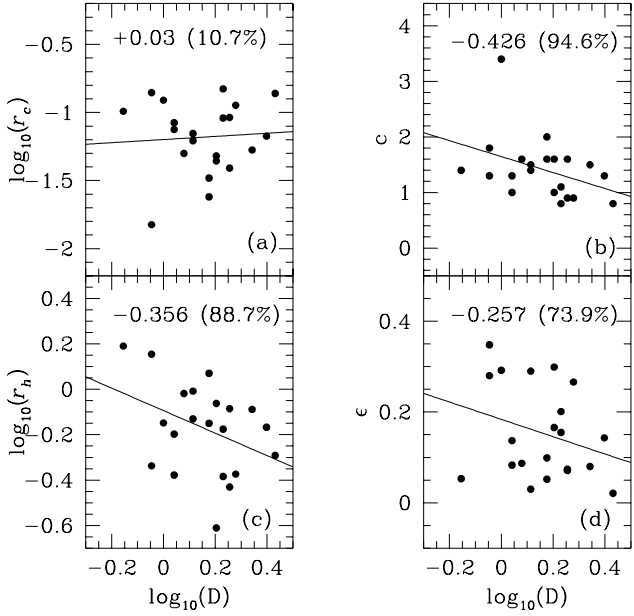}}
\caption{This figure shows the relationships between four properties
of the \object{NGC 5128} GC candidates and their observed distance,
$D$, from the center of \object{NGC 5128}.  The Spearman rank
correlation coefficients, and the confidence levels for rejecting the
hypothesis that the observed correlations are due to chance, are shown
at the top of each panel.  The lines show the best-fitting straight
lines to the data.  Panel (a) shows the core radius (in arcseconds),
panel (b) shows the central concentration, panel (c) shows the
half-mass radius (in arcseconds) of the best-fitting single-mass
Michie--King model, and panel (d) shows the ellipticity.}
\label{FIGURE:distance}
\end{figure}

	GCs live within the tidal field of their parent galaxy, and
the properties of a GC may depend on its position within the
protogalactic cloud at the time the GC formed (e.g.~Murray \&
Lin~\cite{ML92}).  In Fig.~\ref{FIGURE:distance} we plot
$\log_{10}(r_c)$, $\log_{10}(r_h)$ (both radii in arcseconds), $c$,
and $\epsilon$ as functions of $\log_{10}(D)$, where $D$ is the
observed distance from the center of \object{NGC 5128} in arcminutes.
None of the properties show strong correlations with the distance from
the center of \object{NGC 5128}, but the trends are qualitatively
similar to the trends found by Djorgovski \& Meylan~(\cite{DM94}) for
the same properties of the \object{Milky Way}'s GCs (see their
Fig.~8).  The correlations for $\log_{10}(r_c)$ and $c$ for the
\object{NGC 5128} GC candidates are in the same sense as those for the
\object{Milky Way} GCs, although Spearman rank correlation
coefficients suggest that the correlations are weaker for the
\object{NGC 5128} GC candidates.  This is to be expected because we
are using the observed distance from the center of \object{NGC 5128}
(i.e.~the distance projected onto the plane of the sky), while
Djorgovski \& Meylan~(\cite{DM94}) used the true Galactocentric
distance.

	We find a weak trend for the half-mass radius of a GC
candidate to decrease as the distance from the center of \object{NGC
5128} increases, while the opposite trend is seen in the \object{Milky
Way}.  The Spearman rank correlation coefficient for our data is
$-0.356$, which corresponds to a significance of 0.887, while for the
\object{Milky Way} GCs the correlation coefficient is $+0.478$.
Therefore, there is insufficient evidence to for us to conclude that
the observed trend for $r_h$ to decrease as the galactocentric
distance increases is real.

	Djorgovski \& Meylan~(\cite{DM94}) found no correlation
between the ellipticity of a GC and its Galactocentric distance.  We
find a correlation coefficient of $-0.257$, corresponding to a
significance of 0.739.  An examination of Fig.~\ref{FIGURE:distance},
however, suggests that there is no significant correlation between
ellipticity and galactocentric distance for the \object{NGC 5128} GC
candidates.

\section{Colors\label{SECTION:colors}}

\subsection{The Photometry\label{SECTION:photometry}}

        Performing accurate photometry on the \object{NGC 5128} GC
candidates is difficult.  These GC candidates are extended objects,
not point sources, so it is not possible to determine their magnitudes
simply by fitting a scaled PSF to each candidate.  Since each GC
candidate has a unique size \emph{and} radial profile (parameterized
by the concentration, $c$), aperture photometry with a single aperture
will not return accurate magnitudes.  Therefore, we elected to use the
total flux in the best-fitting Michie--King model as the best estimate
of its flux.

        We converted the total instrumental fluxes to the standard
Johnson--Cousins $V\!I$ magnitudes using the prescription of Holtzman
et~al.~(\cite{HB95}).  The calibration equations we used are:

\begin{eqnarray}
\nonumber
V = & -2.5\log_{10}(C_\mathrm{F555W}) + (-0.052 \pm 0.007){(V-I)}_0 \\
\nonumber
    & + (0.027 \pm 0.002){(V-I)}_0^2 \\
    & + (21.725 \pm 0.005) + 2.5\log_{10}(G_i),
\label{EQUATION:calibrate_v}
\end{eqnarray}

\begin{eqnarray}
\nonumber
I = & -2.5\log_{10}(C_\mathrm{F814W}) + (-0.062 \pm 0.009){(V-I)}_0 \\
\nonumber
    & + (0.025 \pm 0.002){(V-I)}_0^2 \\
    & + (20.839 \pm 0.006) + 2.5\log_{10}(G_i),
\label{EQUATION:calibrate_i}
\end{eqnarray}

\noindent
where $C$ is the count rate in ADU/second after correcting for charge
transfer efficiency (CTE) effects. We applied the CTE corrections of
Whitmore \& Heyer~(\cite{WH97}) for a five pixel aperture.  $G$ is the
gain ratio between the 14 e$^-$ gain state (which was used for the
calibration observations) and the 7 e$^-$ gain state (which was used
for the \object{NGC 5128} observations).  As discussed in Holtzman
et~al.~(\cite{HB95}), reddening corrections need to be applied
\emph{before} the instrumental magnitudes are calibrated to the
standard Johnson--Cousins system.  This is because the WFPC2 filters
have different band passes and effective wavelengths than those of the
standard $V$ and $I$ filters.  We adopted a foreground reddening in
the direction of \object{NGC 5128} of $E_{B-V} = 0.11 \pm 0.02$ from
the reddening maps of Burstein \& Heiles~(\cite{BH82}) and assumed
$E_{V\!-\!I} = 1.36E_{B\!-\!V}$ (Taylor~\cite{T86}; Fahlman
et~al.~\cite{FR89}).  We used the extinction corrections for K0III
stars from Table~12 of Holtzman et~al.~(\cite{HB95}) to obtain $A_V =
0.340$ and $A_I = 0.201$.

\begin{figure}
\resizebox{\hsize}{!}{\includegraphics{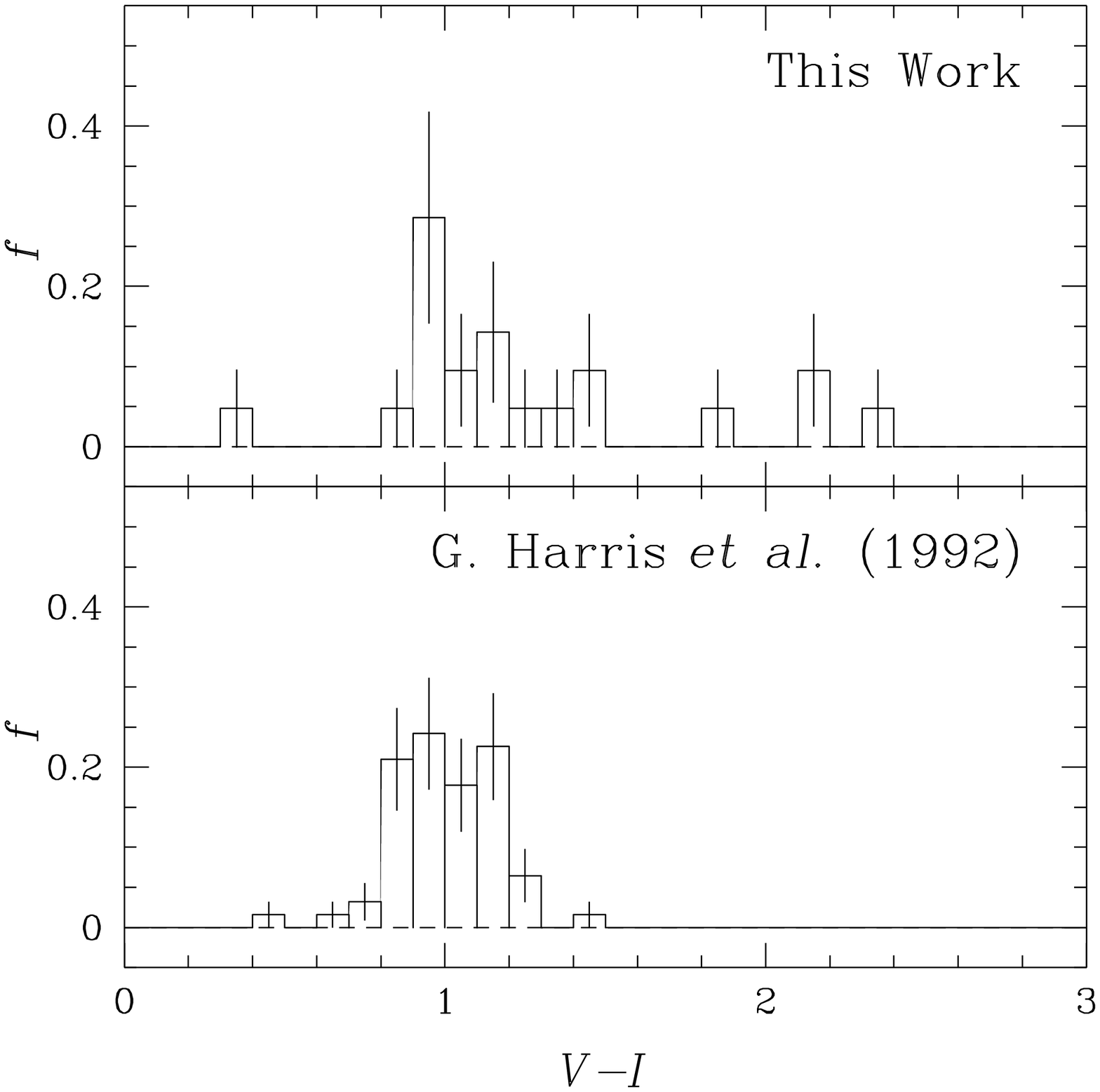}}
\caption{The upper panel shows the distribution of colors for our 21
GC candidates in \object{NGC 5128} while the lower panel shows the
distribution of colors for 62 spectroscopically confirmed GCs in
\object{NGC 5128} from HG92.  The colors have been corrected for
reddening in the \object{Milky Way} in the direction of \object{NGC
5128}, but not for internal reddening in \object{NGC 5128} itself.}
\label{FIGURE:colours}
\end{figure}

        Fig.~\ref{FIGURE:colours} shows the distribution of colors for
our 21 GC candidates, and for 62 spectroscopically confirmed GCs in
\object{NGC 5128} (HG92).  We converted the HG92 $C-T_1$ colors to
$V\!-\!I$ colors using $V\!-\!I = 0.115 + 0.514(C-T_1)$
(Geisler~\cite{G96}).  The $VI$ magnitudes determined this way should
be treated with caution since Geisler~(\cite{G96}) found that $C-T_1$
colors do not reproduce $V\!-\!I$ colors particularly well.  There is
only one object (\#12, $V = 20.30 \pm 0.03$, $V\!-\!I = 0.40 \pm
0.04$) which appears significantly bluer than the other GC candidates.
Based on its color this object may be a young GC\@.  Visual inspection
of the WFPC2 images (see Fig.~\ref{FIGURE:f1_chart}) shows that this
object appears to be a normal GC in \object{NGC 5128}.  The $V$- and
$I$-band photometry of all of our GC candidates are listed in
Table~\ref{TABLE:photometry}.  The $R$-, $J$-, $H$-, and $K$-band
photometry from AM97 are also given.

%
%

\begin{table*}
\begin{center}
\caption{Photometry for the GC candidates in NGC 5128.}
\smallskip
\begin{tabular}{rccccccc}
\hline
\hline
\vspace{0.1 em}
ID & $V$ & $I$ & $V\!-\!I$ & $R$ & $J$ & $H$ & $K$ \\
\hline
\vspace{0.1 em}
  1 & $19.33 \pm 0.02$ & $18.42 \pm 0.02$ & $0.91 \pm 0.03$ & $\cdots$ & $\cdots$ & $\cdots$ & $\cdots$ \\
  2 & $18.21 \pm 0.02$ & $17.22 \pm 0.02$ & $0.99 \pm 0.02$ & $\cdots$ & $\cdots$ & $\cdots$ & $\cdots$ \\
  3 & $21.32 \pm 0.04$ & $19.94 \pm 0.03$ & $1.38 \pm 0.05$ & $\cdots$ & $\cdots$ & $\cdots$ & $\cdots$ \\
  4 & $21.42 \pm 0.04$ & $19.31 \pm 0.03$ & $2.11 \pm 0.05$ & $\cdots$ & $\cdots$ & $\cdots$ & $\cdots$ \\
  5 & $19.76 \pm 0.02$ & $18.49 \pm 0.02$ & $1.27 \pm 0.03$ & $\cdots$ & $\cdots$ & $\cdots$ & $\cdots$ \\
  6 & $19.40 \pm 0.02$ & $17.58 \pm 0.02$ & $1.82 \pm 0.03$ & $\cdots$ & $\cdots$ & $\cdots$ & $\cdots$ \\
  7 & $20.91 \pm 0.03$ & $19.51 \pm 0.03$ & $1.40 \pm 0.04$ & $\cdots$ & $\cdots$ & $\cdots$ & $\cdots$ \\
  8 & $20.08 \pm 0.03$ & $19.11 \pm 0.02$ & $0.97 \pm 0.04$ & $\cdots$ & $\cdots$ & $\cdots$ & $\cdots$ \\
  9 & $19.25 \pm 0.02$ & $18.21 \pm 0.02$ & $1.04 \pm 0.03$ &   19.17  & $17.98 \pm 0.28$ & $17.62 \pm 0.45$ & $\cdots$ \\
 10 & $19.54 \pm 0.02$ & $18.39 \pm 0.02$ & $1.15 \pm 0.03$ &   19.62  & $17.70 \pm 0.19$ & $16.68 \pm 0.15$ & $16.56 \pm 0.26$ \\
 11 & $19.45 \pm 0.02$ & $18.05 \pm 0.02$ & $1.40 \pm 0.03$ & $\cdots$ & $\cdots$ & $\cdots$ & $\cdots$ \\
 12 & $20.30 \pm 0.03$ & $19.90 \pm 0.03$ & $0.40 \pm 0.04$ &   20.51  & $18.56 \pm 0.38$ & $17.74 \pm 0.38$ & $\cdots$ \\
 13 & $18.77 \pm 0.02$ & $17.62 \pm 0.02$ & $1.15 \pm 0.03$ & $\cdots$ & $\cdots$ & $\cdots$ & $\cdots$ \\
 14 & $20.01 \pm 0.02$ & $19.15 \pm 0.02$ & $0.86 \pm 0.03$ &   20.05  & $16.87 \pm 0.16$ & $16.62 \pm 0.30$ & $16.19 \pm 0.25$ \\
 15 & $17.56 \pm 0.02$ & $16.44 \pm 0.02$ & $1.12 \pm 0.03$ &   17.39  & $15.87 \pm 0.07$ & $15.13 \pm 0.06$ & $14.96 \pm 0.08$ \\
 16 & $18.45 \pm 0.02$ & $17.52 \pm 0.02$ & $0.93 \pm 0.03$ &   18.53  & $17.00 \pm 0.12$ & $16.27 \pm 0.12$ & $16.66 \pm 0.24$ \\
 17 & $20.72 \pm 0.03$ & $19.64 \pm 0.03$ & $1.08 \pm 0.04$ & $\cdots$ & $\cdots$ & $\cdots$ & $\cdots$ \\
 18 & $17.07 \pm 0.02$ & $16.08 \pm 0.02$ & $0.99 \pm 0.03$ &   17.09  & $15.49 \pm 0.05$ & $14.93 \pm 0.04$ & $14.58 \pm 0.05$ \\
 19 & $19.97 \pm 0.03$ & $17.78 \pm 0.03$ & $2.19 \pm 0.04$ & $\cdots$ & $\cdots$ & $\cdots$ & $\cdots$ \\
 20 & $20.03 \pm 0.03$ & $17.65 \pm 0.03$ & $2.38 \pm 0.04$ & $\cdots$ & $\cdots$ & $\cdots$ & $\cdots$ \\
 21 & $18.41 \pm 0.02$ & $17.41 \pm 0.02$ & $1.00 \pm 0.03$ & $\cdots$ \\
\hline
\hline
\end{tabular}
\label{TABLE:photometry}
\end{center}
\end{table*}

\subsection{Reddening Within NGC 5128\label{SECTION:reddening}}

        In order to estimate the amount of internal reddening within
\object{NGC 5128} in the direction of each GC candidate we determined
the color of the background near each object.  This was done by
letting the mean background color be a free parameter during the
Michie--King model fitting process.  The expected unreddened $V\!-\!I$
color at the location of each GC candidate was then subtracted from
the observed mean background color to get an estimate of the internal
reddening.  The expected color of the background that each GC
candidate sits on was determined as follows.

	From Fig.~5 of van~den~Bergh~(\cite{V76}) we derived

\begin{equation}
B\!-\!V = 0.153\log_{10}(D) + 0.924,
\label{EQUATION:BV_gradient}
\end{equation}

\noindent
where $D$ is the projected distance from the center of \object{NGC
5128} in arcminutes.  We estimate that the uncertainty in the
$B\!-\!V$ values from Eq.~\ref{EQUATION:BV_gradient} is $\sim \pm
0.1$.  The $B\!-\!V$ colors were converted to ${(V\!-\!I)}_0$ colors
by subtracting the adopted \object{Milky Way} reddening of
$E_{B\!-\!V} = 0.11 \pm 0.02$ and using

\begin{equation}
{(V\!-\!I)}_0 = 0.745{(B\!-\!V)}_0 + 0.399,
\label{EQUATION:VI_BV_relation}
\end{equation}

\noindent
which we derived from the Galactic Globular Cluster Catalogue of
W. Harris~(\cite{H96}).  For each GC candidate in \object{NGC 5128} we
computed the expected background color using
Eqs.~\ref{EQUATION:BV_gradient} and~\ref{EQUATION:VI_BV_relation} and
subtracted this from the mean color of the unresolved background
around each GC candidate.  The internal reddening due to dust in
\object{NGC 5128} was then computed by subtracting the expected color
from the observed color for each GC candidate.  The internal
reddenings, as well as the dereddened $V_0$ and ${(V\!-\!I)}_0$ values
for each GC candidate, are listed in Table~\ref{TABLE:reddening}.  The
$V_0$ and ${(V\!-\!I)}_0$ values are corrected for both the Burstein
\& Heiles~(\cite{BH82}) reddening \emph{and} our estimate of the
internal reddening in \object{NGC 5128}.  The uncertainty in each
reddening estimate is $\sim 0.08$ and negative internal reddenings
were set to $E_{V\!-\!I} = 0$.


%
%

\begin{table}
\begin{center}
\caption[]{The estimated internal reddening within \object{NGC 5128}
along the line of sight, the dereddened $V$-band magnitude, and the
dereddened ${(V\!-\!I)}_0$ color for each GC candidate.}
\smallskip
\begin{tabular}{rccc}
\hline
\hline
\vspace{0.1 em}
ID & $E_{V\!-\!I}$ & $V_0 \pm \sigma_{V_0}$ & ${(V\!-\!I)}_0 \pm \sigma_{{(V\!-\!I)}_0}$ \\
\hline
\vspace{0.1 em}
  1 & 0.17 & $18.94 \pm 0.06$ & $0.74 \pm 0.08$ \\
  2 & 0.15 & $17.87 \pm 0.06$ & $0.84 \pm 0.08$ \\
  3 & 0.06 & $21.18 \pm 0.07$ & $1.31 \pm 0.09$ \\
  4 & 0.00 & $21.42 \pm 0.07$ & $2.12 \pm 0.10$ \\
  5 & 0.00 & $19.76 \pm 0.06$ & $1.27 \pm 0.09$ \\
  6 & 0.07 & $19.25 \pm 0.06$ & $1.75 \pm 0.09$ \\
  7 & 0.14 & $20.59 \pm 0.07$ & $1.27 \pm 0.09$ \\
  8 & 0.06 & $19.94 \pm 0.06$ & $0.90 \pm 0.09$ \\
  9 & 0.08 & $19.08 \pm 0.06$ & $0.96 \pm 0.08$ \\
 10 & 0.11 & $19.28 \pm 0.06$ & $1.03 \pm 0.09$ \\
 11 & 0.09 & $19.26 \pm 0.06$ & $1.32 \pm 0.09$ \\
 12 & 0.14 & $19.98 \pm 0.06$ & $0.26 \pm 0.09$ \\
 13 & 0.10 & $18.53 \pm 0.06$ & $1.04 \pm 0.08$ \\
 14 & 0.10 & $19.77 \pm 0.06$ & $0.76 \pm 0.09$ \\
 15 & 0.06 & $17.42 \pm 0.06$ & $1.06 \pm 0.08$ \\
 16 & 0.14 & $18.13 \pm 0.06$ & $0.79 \pm 0.08$ \\
 17 & 0.13 & $20.42 \pm 0.07$ & $0.95 \pm 0.09$ \\
 18 & 0.10 & $16.83 \pm 0.06$ & $0.89 \pm 0.08$ \\
 19 & 0.10 & $19.76 \pm 0.07$ & $2.10 \pm 0.09$ \\
 20 & 0.00 & $20.03 \pm 0.07$ & $2.38 \pm 0.09$ \\
 21 & 0.22 & $17.91 \pm 0.06$ & $0.78 \pm 0.08$ \\
\hline
\hline
\end{tabular}
\label{TABLE:reddening}
\end{center}
\end{table}

	In order to check these differential reddenings, we repeated
our calculation using the color of a typical galaxy with the same
morphological type as \object{NGC 5128}.  The Third Reference
Catalogue of Bright Galaxies (de~Vaucouleurs et~al~\cite{dV91}) lists
\object{NGC 5128} as being an intermediate S0 galaxy with a
morphological type of $T = -2 \pm 0.3$.  Buta \&
Williams~(\cite{BW95}) find a mean color for $T = -2$ galaxies of
$\langle V\!-\!I \rangle_\mathrm{total} = 1.145 \pm 0.069$ based on a
sample of 55 galaxies.  This is within $1\sigma$ of the expected
background colors derived from the observed color gradient in the
outer regions of \object{NGC 5128}.  If we assume that the unresolved
light from \object{NGC 5128} is the same as that from a typical $T =
-2$ galaxy, but has been reddened due to the presence of dust from the
merger, then the excess reddening can be estimated by subtracting the
mean color of a $T = -2$ galaxy from the observed color of the
unresolved background around each GC candidate.  The differential
reddenings that we obtain by assuming that \object{NGC 5128} is a $T =
-2$ galaxy are consistent with those obtained using the
van~den~Bergh~(\cite{V76}) color gradient.

        The first problem with this method of estimating the internal
reddening in \object{NGC 5128} is that \object{NGC 5128} is not a
normal elliptical galaxy, but an elliptical galaxy that has undergone
a merger with a small late-type spiral galaxy.  Therefore, the
unresolved light near the center will be a combination of light from
the original galaxy and from stars in the captured spiral galaxy.  In
our derivations of the internal reddening we have assumed that the
only changes in the color of the central regions of \object{NGC 5128}
are those due to dust, and ignored changes in color due to star
formation and differences in the underlying stellar population.  Buta
\& Williams~(\cite{BW95}) list mean colors for late spiral galaxies of
$\langle V\!-\!I \rangle_\mathrm{total} \sim 0.8$ to $0.7$, depending
on the morphological type of the spiral.  Therefore, the contribution
from the merged spiral galaxy will cause the actual unreddened color
of the unresolved background in \object{NGC 5128} to be as much as a
few tenths of a magnitude bluer than we have assumed.  This will lead
us to underestimate the internal reddening within \object{NGC 5128} by
a few tenths of a magnitude.  It is possible that some of the GC
candidates are being seen against regions of recent star formation in
\object{NGC 5128}.  This would also lead us to underestimate the
internal reddening by a few tenths of a magnitude.
              
        A second problem is that the structure of the dust features in
\object{NGC 5128} changes on spatial scales of less than $\sim
1\arcsec$.  The mean tidal radius of the GC candidates is
$\overline{r_t} = 1\farcs42 \pm 0\farcs15$ (se), and the surface
brightness of the background was determined while fitting the
Michie--King models by taking the mean value of the pixels that fell
within the fitting box ($64 \times 64$ pixels) but beyond the tidal
radius.  Therefore, small-scale spatial structure in the dust lane
could introduce errors of several tenths of a magnitude in our
estimates of the internal reddening in \object{NGC 5128}.

        Finally, the reddening corrections are not being made in the
WFPC2 photometric system.  This may introduce small errors in the
reddening corrections that we determine.  We believe, however, that
these errors will be small compared to other uncertainties in the
method.  In light of these uncertainties we believe that it is
possible that our estimates of the differential reddening for each GC
candidate are only accurate to $\sim \pm 0.3$ mag.

\begin{figure}
\resizebox{\hsize}{!}{\includegraphics{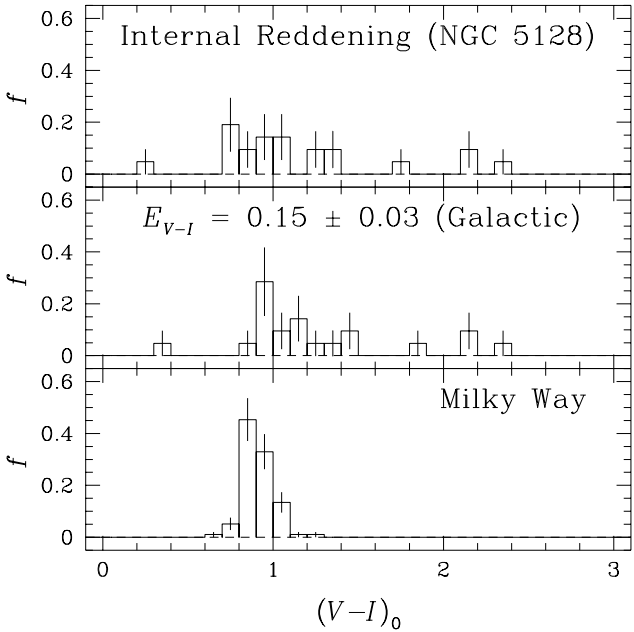}}
\caption{The upper panel shows the distribution of colors for the GC
candidates in \object{NGC 5128} after correcting for the estimated
internal reddening due to dust in \object{NGC 5128}.  The middle panel
shows the distribution of colors for the GC candidates without any
correction for internal reddening within \object{NGC 5128} (but with a
correction for Galactic reddening in the direction of \object{NGC
5128}).  The lower panel shows the distribution of dereddened colors
for the \object{Milky Way} GC system.}
\label{FIGURE:reddening}
\end{figure}

        Fig.~\ref{FIGURE:reddening} shows the color distribution of
the GC candidates after correcting for internal reddening within
\object{NGC 5128}.  There appear to be three populations in the upper
panel of Fig.~\ref{FIGURE:reddening}.  The largest population is
centered at ${(V\!-\!I)}_0 \sim 1.0$ and contains 16 of the 21 GC
candidates.  This population appears to be similar to the
spectroscopically confirmed GCs of HG92 and probably represents a
genuine old population of GCs in \object{NGC 5128}.  The similarity
between the colors of these GC candidates and the dereddened colors of
the \object{Milky Way} GCs suggests that this population is not
heavily reddened and therefore probably lies on the near side of
\object{NGC 5128}.  The second component is the four objects with
${(V\!-\!I)}_0 > 1.4$.  These objects are probably GCs that are being
seen through large amounts of dust in \object{NGC 5128}.  Alternately,
they could be background galaxies that have been misidentified as GC
candidates.  The third component is the single blue object with
${(V\!-\!I)}_0 = 0.26 \pm 0.09$.  An examination of the image of this
object (Fig.~\ref{FIGURE:f1_chart}) suggests that it is a legitimate
GC candidate so its blue color makes it the best candidate for being a
young GC in our sample.

\subsection{Iron Abundance\label{SECTION:metallicity}}

	The iron abundance for each GC candidate was estimated using

\begin{equation}
\mathrm{[Fe/H]} = -6.568 + 6.733{(V\!-\!I)}_0 - 1.372{(V\!-\!I)}_0^2,
\label{EQUATION:feh_relation}
\end{equation}

\noindent
which was determined from 45 Galactic GCs with low reddenings and good
metallicity estimates, and from 13 metal-rich GCs in NGC 1399
(Kissler-Patig et~al.~\cite{KP98}).

	If the blue color of object \#12 is due solely to its iron
abundance, then it has $\mathrm{[Fe/H]} = -4.91 \pm 0.54$.  This
implausibly low iron abundance suggests that age is primarily
responsible for at least some of the blue color, supporting the notion
that \#12 is a young object.  The mean iron abundance of the 16 GC
candidates with $0.6 \le {(V\!-\!I)}_0 \le 1.4$ is
$\overline{\mathrm{[Fe/H]}} = -1.27 \pm 0.20$ (se), which is within
$2\sigma$ of the $\mathrm{[Fe/H]} = -0.8 \pm 0.2$ value found by HG92.
This suggests that the colors of these objects are consistent with
them being old GCs, or moderately reddened intermediate-age GCs.  The
four objects with ${(V\!-\!I)}_0 > 1.4$ all have iron abundances of
greater than the Solar value ($\overline{\mathrm{[Fe/H]}} = +1.44 \pm
0.15$).  This suggests that the red colors of these objects are
primarily due to dust along the line of sight within \object{NGC
5128}, not high iron abundances, although we can not rule out the
possibility that at least some of these GC candidates have high iron
abundances.


\section{Young GCs in NGC 5128\label{SECTION:young_GCs}}

	We would expect young GCs to be brighter and bluer than their
old counterparts, although reddening uncertainties complicate the
interpretation of the observed colors for the \object{NGC 5128} GC
candidates.  The mean ${(V\!-\!I)}_0$ color of the 97 \object{Milky
Way} GCs that have both $V\!-\!I$ and $E_{B\!-\!V}$ values listed in
the catalogue of W. Harris~(\cite{H96}) is $0.91 \pm 0.09$ (standard
deviation).  This is consistent with the colors of the majority of the
GC candidates that we found in the inner regions of \object{NGC 5128}.

\begin{figure}
\resizebox{\hsize}{!}{\includegraphics{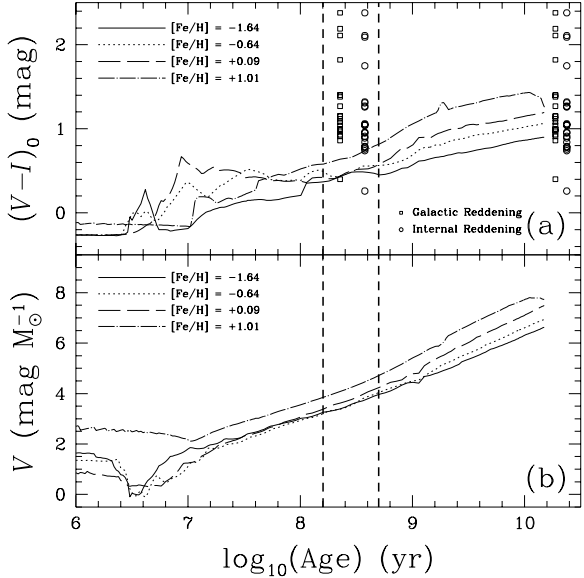}}
\caption{The upper panel shows the time evolution of ${(V\!-\!I)}_0$
for GCs with a Salpeter Initial Mass Function for a range of cluster
masses from 0.1 to 125 Solar masses, and a range of iron abundances
from $\mathrm{[Fe/H]} = -1.64$ to $\mathrm{[Fe/H]} = +1.01$, using the
models of Bruzual \& Charlot~(\protect\cite{BC93}).  The heavy
vertical lines are at 160 Myr and 500 Myr respectively (the interval
that the most recent merger most likely occurred in).  The colors of
the 21 GC candidates are plotted as squares (Galactic reddening
removed) and circles (Galactic \emph{and} internal reddening within
\object{NGC 5128} removed).  The lower panel shows the time evolution
of the $V$ magnitude per Solar mass using the same models.}
\label{FIGURE:ages}
\end{figure}

        Schweizer \& Seitzer~(\cite{SS93}) find that young GCs in
\object{NGC 7252} have $V\!-\!I$ colors that are $\sim 0.5$ mag bluer
than old GCs, which suggests that objects in \object{NGC 5128} with
colors bluer than $V\!-\!I \sim 0.8$ may be young ($t_0 < 1$ Gyr) GCs.
We find five objects with ${(V\!-\!I)}_0 < 0.8$.  One of these, object
\#12, is unusually blue, with ${(V\!-\!I)}_0 = 0.26 \pm 0.09$.  AM97
identified this object as M23 and measured a $J\!-\!H$ color of 0.83.
A visual examination of the \emph{HST} images (see
Fig.~\ref{FIGURE:f1_chart}) shows nothing unusual about this object.
If we assume that the blue color of this object is due to it being a
young star cluster, we can use the models of Bruzual \&
Charlot~(\cite{BC93}) to estimate its age and mass from its
${(V\!-\!I)}_0$ color and its integrated $V$ magnitude.
Fig.~\ref{FIGURE:ages}a shows the evolution of the ${(V\!-\!I)}_0$
color of star clusters for several iron abundances.  This figure
suggests an age for \#12 of less than $\sim 100$ Myr, while
Fig.~\ref{FIGURE:ages}b suggests that, if it is a young, metal-poor
GC, it will fade by between $\sim 4$ and $\sim 6.5$ mag over the next
12 Gyr.  Therefore, when object \#12 is as old as a typical
\object{Milky Way} GC ($\sim 12$ Gyr), it will have $M_V$ between
$\sim -3.5$ and $-1$, corresponding to a mass of between $\sim 4000$
and $\sim 400$ Solar masses, assuming a mass-to-light ratio of two.
W. Harris~(\cite{H96}) lists four Galactic GCs with masses less than
4000 Solar masses: \object{AM 1} with a mass of 700 Solar masses,
\object{Pal 1} with a mass of 1000 Solar masses, \object{E 3} with a
mass of 2100 Solar masses, and \object{Terzan 1} with a mass of 3400
Solar masses, although these mass determinations are highly uncertain.
In light of the small mass for object \#12 it is also possible that
this object is a massive open cluster.

	The sixteen GC candidates with $0.7 \le {(V\!-\!I)}_0 < 1.4$
have colors that are consistent with being young ($< 1$ Gyr),
intermediate-aged (1--4 Gyr), or old ($> 4$ Gyr) GCs depending on the
iron abundances and reddenings that are assumed.  If these GC
candidates have iron abundances similar to those found in the Galactic
GCs ($\mathrm{[Fe/H]} < 0$) then Fig.~\ref{FIGURE:ages}a suggests that
their observed ${(V\!-\!I)}_0$ colors are consistent with them having
ages of $> 1$ Gyr.  If this is the case then Fig.~\ref{FIGURE:ages}b
suggests that these objects will fade by between $\sim 0$ and $2$ mag
in $V$ by the time they are $\sim 12$ Gyr old.  This corresponds to
these GC candidates having masses between $\sim 40\,000$ and $\sim
4\,000\,000$ Solar masses, which is consistent with the masses of
Galactic GCs.  If, on the other hand, the objects have iron abundances
greater than the Solar value then their colors are consistent with
their being young GCs.  In this case Fig.~\ref{FIGURE:ages}b suggests
that they will fade by $\sim 3$ to $4$ mag in $V$ over then next 12
Gyr, which would give them masses of between $\sim 5000$ and $\sim
250\,000$ Solar masses.

	These results assume that the differential reddenings derived
in Sect.~\ref{SECTION:reddening} are uncertain by $\sim \pm 0.3$ mag,
and may contain systematic uncertainties of a few tenths of a
magnitude.  The ${(V\!-\!I)}_0$ colors of the sixteen GC candidates
with $0.7 \le {(V\!-\!I)}_0 < 1.4$ are most consistent with young GCs
if the differential reddenings have been systematically underestimated
by between 0.1 and 0.7 mag.  If our estimates of the differential
reddening for these objects are correct then their ${(V\!-\!I)}_0$
colors suggest that these GC candidates are normal, old, metal-poor
GCs.  Spectroscopic observations are needed to break the degeneracy
between reddening and metallicity to accurately determine the ages of
these objects.

	The red colors, and high implied iron abundances, of the four
GC candidates with ${(V\!-\!I)}_0 \ge 1.4$ suggest that these objects
are heavily reddened.  If we assume that all of these objects are
young, metal rich clusters that formed at the time of the merger event
then differential reddenings of $E_{V\!-\!I} \sim 1.1$ to 1.8,
\emph{in addition to those derived in Sect.~\ref{SECTION:reddening}},
are required to account for their observed red colors.  This
corresponds to between $\sim 2.4$ and $\sim 3.9$ mag of extra
extinction in the $V$ band.  If, on the other hand, these objects are
young metal poor clusters then the amount of \emph{additional}
differential reddening required to account for their observed red
colors is between $\sim 1.3$ and $\sim 2.1$ mag.  It is unlikely that
we have underestimated the differential reddening by this amount.  If
we assume that these objects are old GCs ($\sim 12$ Gyr) seen through
dust then we have underestimated the amount of differential reddening
by between $\sim 0.4$ and $\sim 1.4$ mag.  It is possible that a
combination of random and systematic uncertainties in the differential
reddening calculations could result in us underestimating the
reddenings of these objects by several tenths of a magnitude, so we
can not rule out the possibility that these four objects are old GCs.
Another possibility is that they are background red elliptical, or
dwarf elliptical galaxies with concentrations similar to those of the
Galactic GCs.  Spectroscopic observations are needed to confirm the
nature of these objects.

	In summary, the dereddened ${(V\!-\!I)}_0$ color of object
\#12 is consistent with it being either a small young GC or a massive
open cluster.  The sixteen objects with $0.8 \le {(V\!-\!I)}_0 < 1.4$
have colors that are consistent with them being either young GCs which
suffer from up to $\sim 0.7$ mag of differential reddening or old GCs
that suffer from up to $\sim 0.2$ mag of differential reddening.
Spectroscopic observations will be needed to determine the ages of
these GC candidates.  The four objects with ${(V\!-\!I)}_0 > 1.4$ are
either old heavily reddened GCs or background galaxies that have
similar concentrations to Galactic GCs.


\section{Conclusions\label{SECTION:conc}}

	We have used the \emph{HST}/WFPC2 to identify 21 GC candidates
in a $25 \sq\arcmin$ area centered on the nucleus of the nearest giant
elliptical galaxy, \object{NGC 5128}.  There is strong evidence that
this galaxy has undergone a merger with a small late-type spiral
between $\sim 160$ and $360$ Myr ago.  We fit two-dimensional,
PSF-convolved, single-mass Michie--King models to each GC candidate
and derived core radii, tidal radii, half-mass radii, ellipticities,
and position angles for each object.  We assumed that the GC
candidates were structurally similar to those in the Local Group;
i.e., they could be well fit by Michie--King models, and had similar
half-mass radii and ellipticities.  Therefore, only objects with
structural parameters similar to those of GCs in the Local Group were
accepted as being GC candidates.  It is possible that we have missed
\object{NGC 5128} GC candidates that are significantly less centrally
concentrated than the GCs in the Local Group.  However, since we were
primarily interested in using the colors of the GC candidates to
estimated their ages we preferred to risk rejecting legitimate GC
candidates rather than risk having our sample contaminated with stars
and background galaxies.  Within this constraint we find no evidence
that the \object{NGC 5128} GC candidates have core-, half-mass-, or
tidal radii that are distributed differently from those GCs in the
\object{Milky Way} that do not exhibit central brightness cusps.
There is weak evidence that \object{NGC 5128} GC candidates are
systematically more elliptical than are the Galactic GCs.

	We have obtained $V$- and $I$-band photometry for all 21 of
our GC candidates.  We find no evidence for the bimodal color
distribution observed among GCs at larger distances from the center of
\object{NGC 5128} (HG92; Zepf \& Ashman~\cite{ZA93}), although this is
likely due to the small sample size (21 objects), the poor metallicity
sensitivity of the $V\!-\!I$ color index, and confusion due to
differential reddening within \object{NGC 5128}.

	We have identified one very blue GC candidate (\#12) with
${(V\!-\!I)}_0 = 0.26 \pm 0.09$.  Using a color-age relation derived
from the models of Bruzual \& Charlot~(\cite{BC93}), we estimate that
this object has an age of less than $\sim 100$ Myr and a mass of
between $\sim 400$ and $\sim 4000$ Solar masses, which is barely
consistent with this object being a small GC that formed during the
merger event.  There are sixteen objects with $0.7 \le {(V\!-\!I)}_0 <
1.4$ that have colors and integrated magnitudes that are consistent
with their being either young GCs that formed during the merger event
or old GCs similar to those found in the \object{Milky Way}, depending
on what we assume about their iron abundances and differential
reddenings.  The colors of these GCs are very similar to those
measured by HG92, but the amount of reddening for each object is
uncertain by $\sim 0.3$ mag.  There are four GC candidates with
${(V\!-\!I)}_0 > 1.4$, which imply either implausibly high
metallicities or very large differential reddenings.  We can not rule
out the possibility that some of these four objects are background
galaxies seen through the central regions of \object{NGC 5128}.

	Spectroscopic studies will be needed to determine
unambiguously if any of the central GC candidates in \object{NGC 5128}
are young GCs that may have formed as a result of the merger.  If
young GCs in \object{NGC 5128} can be unambiguously identified
spectroscopically, and their ages determined, this will make it
possible to estimate the amount of time required for GCs to form after
a galactic merger has occurred.


\begin{acknowledgements}

       This research is based on observations made with the NASA/ESA
\emph{Hubble Space Telescope} obtained at the Space Telescope Science
Institute.  STScI is operated by the Association of Universities for
Research in Astronomy Inc.\ under NASA contract NAS 5-26555.  S. H. is
supported by the Danish Centre for Astrophysics with the \emph{HST}.
P. C.  acknowledges financial support from the Sherman M. Fairchild
Foundation.  The authors would like to thank Peter Stetson for kindly
making his WFPC2 PSFs, and the {\sc daophot ii} software, available.

\end{acknowledgements}




\end{document}